\documentclass[useAMS,usenatbib,fleqn]{mn2e} % comment this line when submitting

\pdfoutput=1

\usepackage{amsmath}
\usepackage{graphicx}
\usepackage[german,british]{babel}
\usepackage[varg]{txfonts}
\usepackage{biblio}
\usepackage{natbib}
\usepackage{color}
\usepackage{microtype}

\usepackage[draft]{hyperref}

\hypersetup{%
  colorlinks=true,
  linkcolor=red,
  citecolor=blue,
  bookmarksopen=true,
  bookmarksnumbered=true,
  pdfstartpage={1},  
  pdftitle = {},
  pdfsubject = {},
  pdfauthor = {} ,
  pdfkeywords = {},
  pdfproducer = {LaTeX with hyperref}
}

\providecommand{\eprint}[1]{\href{http://arxiv.org/abs/#1}{#1}}

\newlength{\colwidth}
\newcommand{\cm}{{\rm cm}}
\newcommand{\psc}{{\rm cm^{-2}}}
\newcommand{\pcc}{{\rm cm^{-3}}}

\newcommand{\kms}{{\, \rm km}\,{\rm s}^{-1}}

\newcommand{\cms}{{\, \rm cm}\,{\rm s}^{-1}}
\newcommand{\K}{{\, \rm K}}

\newcommand{\hkpc}{h^{-1}\,{\rm kpc}}
\newcommand{\kpc}{{\rm kpc}}
\newcommand{\Mpc}{{\rm Mpc}}

\newcommand{\hMpc}{h^{-1}\,{\rm Mpc}}

\newcommand{\hMsun}{{h^{-1}\,{\rm M}_\odot}}
\newcommand{\Msun}{{{\rm M}_\odot}}

\newcommand{\ion}[2]{\hbox{#1\,{\sc #2}}}
\newcommand{\ionsubscript}[2]{\hbox{\scriptsize #1\,{\tiny #2}}}
\renewcommand{\H}{{\rm H}}

\newcommand{\N}{{\rm N}}
\newcommand{\Ox}{{\rm O}}
\newcommand{\Ne}{{\rm Ne}}

\newcommand{\HI}{\ion{H}{i}}

\newcommand{\CIII}{\ion{C}{iii}}
\newcommand{\CIV}{\ion{C}{iv}}

\newcommand{\NII}{\ion{N}{ii}}
\newcommand{\NIII}{\ion{N}{iii}}
\newcommand{\NIV}{\ion{N}{iv}}
\newcommand{\NV}{\ion{N}{v}}

\newcommand{\NeVIII}{\ion{Ne}{viii}}
\newcommand{\NeVII}{\ion{Ne}{vii}}

\newcommand{\OIII}{\ion{O}{iii}}
\newcommand{\OIV}{\ion{O}{iv}}

\newcommand{\OVI}{\ion{O}{vi}}

\newcommand{\SiIII}{\ion{Si}{iii}}

\newcommand{\MgII}{\ion{Mg}{ii}}
\newcommand{\MgX}{\ion{Mg}{x}}

\newcommand{\SuIII}{\ion{S}{iii}}
\newcommand{\SuIV}{\ion{S}{iv}}
\newcommand{\SuV}{\ion{S}{v}}

\newcommand{\lya}{Ly$\alpha$}

\newcommand{\bea}{\begin{eqnarray}}
\newcommand{\eea}{\end{eqnarray}}
\newcommand{\beq}{\begin{equation}}
\newcommand{\eeq}{\end{equation}}
\newcommand{\bit}{\begin{itemize}}
\newcommand{\eit}{\end{itemize}}
\newcommand{\ben}{\begin{enumerate}}
\newcommand{\een}{\end{enumerate}}
\newcommand{\los}{sightline}
\newcommand{\loss}{sightlines}

\newcommand{\NeVIIIstrong}{\mbox{770.41~\AA}}
\newcommand{\NeVIIIweak}{\mbox{780.32~\AA}}

\newcommand{\NOVI}{{\rm N_{\ionsubscript{O}{VI}}}}

\newcommand{\NNeVIII}{{\rm N_{\ionsubscript{Ne}{VIII}}}}

\newcommand{\NHI}{{N_{\ionsubscript{H}{I}}}}

\newcommand{\nHI}{{n_{\ionsubscript{H}{I}}}}

\newcommand{\nH}{{{n}_{\rm H}}}

\newcommand{\nNe}{{{n}_{\rm Ne}}}

\newcommand{\nNeVIII}{{n_{\ionsubscript{Ne}{VIII}}}}

\newcommand{\NH}{{N_{\rm H}}}

\newcommand{\fHI}{{f_{\ionsubscript{H}{I}}}}

\newcommand{\Xne}{\left( \nNe / \nH \right)}

\newcommand{\fNeVIII}{{f_{\ionsubscript{Ne}{VIII}}}}

\newcommand{\fhi}{(\nHI/ \nH)}

\newcommand{\fneviii}{(\nNeVIII/\nNe)}

\newcommand{\bneviii}{b_{\ionsubscript{Ne}{VIII}}}

\newcommand{\bhi}{b_{\ionsubscript{H}{I}}}

\title[ \NeVIII\ absorbers at low redshift]{ Absorption signatures of warm-hot gas at low redshift: \NeVIII}

\author[Tepper-Garc\'\i{}a et al.]{%
Thor Tepper-Garc\'\i{}a,$^{1}$\thanks{E-mail: tepper@astro.physik.uni-potsdam.de}
Philipp Richter,$^{1,2}$ 
and Joop Schaye$^{3}$ 
\\
$^{1}$Institut f\"ur Physik und Astronomy, Universit\"at Potsdam, Karl-Liebknecht-Str. 24/25, 14476 Potsdam, Germany\\
$^{2}$Leibniz-Institut f\"ur Astrophysik Potsdam (AIP), An der Sternwarte 16, 14482 Potsdam, Germany\\
$^{3}$Leiden Observatory, Leiden University, P.O. Box 9513, 2300 RA Leiden, The Netherlands\\
}
\begin{document}

\date{Accepted ----. Received ----; in original form ----}

\pagerange{\pageref{firstpage}--\pageref{lastpage}} \pubyear{----}

\maketitle

\label{firstpage}

\begin{abstract}
At \mbox{$z<1$} a large fraction of the baryons is thought to reside in diffuse gas that has been shock-heated to high temperatures ($10^5-10^6~\K$). Absorption by the 770.41, 780.32 \AA\ doublet of NeVIII in quasar spectra represents a unique tool to study this elusive warm-hot phase. We have developed an analytic model for the properties of NeVIII absorbers that allows for an inhomogeneous metal distribution. Our model agrees with the predictions of a simulation from the OWLS project indicating that the average line-of-sight metal-filling fraction within the absorbing gas is low ($c_L\sim0.1$). Most of the NeVIII in our model is produced in low-density, collisionally ionised gas (\mbox{$\nH=10^{-6}-10^{-4}~\pcc$}, \mbox{$T=10^5-10^6~\K$}). Strong NeVIII absorbers ($\log_{10}(\NNeVIII/\psc)\gtrsim14$), like those recently detected by HST/COS, are found to arise in higher density gas ($\nH\gtrsim10^{-4}\pcc$, $T\approx5\times10^5~\K$). NeVIII cloudlets harbour only 1 per cent of the cosmic baryon budget. The baryon content of the surrounding gas (which has similar densities and temperatures as the NeVIII cloudlets) is a factor $c_L^{-1}$ higher. We conclude that NeVIII absorbers are robust probes of shock-heated diffuse gas, but that spectra with signal-to-noise ratios \mbox{${\rm~S/N}>100$} would be required to detect the bulk of the baryons in warm-hot gas.
\end{abstract}


%--------------------------------------------------------------------------------------------------------------------------------------------------------------------------------
\section{Introduction} \label{sec:intro}

The analysis of the physical conditions and the distribution and abundances of heavy elements in interstellar- and intergalactic gas using absorption spectroscopy is a well-established technique which has yielded a wealth of information about the status and the evolution of a significant fraction of the baryonic matter in the Universe. The {\em Cosmic Origins Spectrograph} \citep[COS;][]{gre98a, gre12a}, an ultra-violet (UV) spectrograph installed on the {\em Hubble Space Telescope} (HST), has opened a new era of absorption spectroscopy, as a result of its increased sensitivity (from factors of a few to an order of magnitude) with respect to previous instruments such as the {\em Far Ultraviolet Spectroscopic Explorer} (FUSE) satellite or the {\em Space Telescope Imaging Spectrograph} (STIS) installed on HST. COS is designed to perform high-sensitivity, medium- and low-resolution (FWHM~$\approx 17 \kms$) spectroscopy of astronomical objects in the 1150~\AA\ -- 3200~\AA\ wavelength range.

One of the major science goals of COS is to clarify the long standing mystery of the whereabouts of  a significant fraction (30 to 50 per cent) of the baryons in the low-redshift Universe \citep{per92a,fuk04a,shu11a}, that were synthesised in the Big Bang and that are detected mainly through \HI\ absorption at redshift $z \gtrsim 3$ \citep{rau97a,wei97b,sch01a}. A robust prediction of cosmological simulations is that most of these ``missing baryons'' reside in a gas phase with densities $\nH \sim 10^{-6} - 10^{-5} \pcc$ that, as a result of the formation of structure in the Universe, has been shock-heated to temperatures $T \gtrsim 10^5 \K$ \citep[e.g.][]{cen99a,dav01a,ber08a}. Along with gravitational shock-heating, feedback processes such as supernova (SN) explosions and outflows driven by active galactic nuclei (AGN) have been shown to significantly contribute to the mass  in this diffuse, warm-hot gas phase at low redshift \citep[e.g.][]{tep12a}.

So far, attempts to trace shock-heated diffuse gas at low redshift, and to constrain its baryon content, have mainly relied on the detection of absorption by five-times ionised oxygen (\OVI) in the UV spectra of distant point-like sources \citep[e.g.\ quasars, QSOs; ][]{tri00a,ric04a,dan06a,dan08b,tho08a,tri08b,dan10a,pro11a}. However, because \OVI\ may be produced both by photo-ionisation in gas at $T \sim 10^4 \K$ and by collisional ionisation in gas at $T\approx 3\times10^5 \K$, and because the metals may well be poorly mixed on small scales \citep{sch07a}, the interpretation of intervening \OVI\ absorbers and their status as tracers of shock-heated gas is still controversial, as shown by both observational \citep{tho08b,tri08b,dan08b} and theoretical studies \citep[e.g.][]{opp09b,tep11a,smi11a,cen12a}. Besides \OVI, broad ($\bhi \geq 40 \kms$) and shallow ($\tau_0(\HI) \lesssim 0.1$) \lya\ absorption features, the so-called Broad \lya\ Absorbers \citep[BLAs;][]{ric06a}, have been considered as a means to detect diffuse warm-hot gas at low redshift \citep{sem04a,ric04a,wil06a,leh07a,dan10a,tep12a}. However, BLAs are difficult to identify in the UV spectra of QSOs because of the limited signal-to-noise (S/N) and spectral resolution of data obtained with current space-based UV spectrographs.

An alternative to \OVI\ and BLAs as tracers of shock-heated gas might be offered by seven-times ionised neon (\NeVIII). Neon is the fifth most abundant element in the Universe,\footnote{After hydrogen, helium, oxygen and carbon.} and this particular ionisation state has two resonant transitions with rest-frame wavelengths in the extreme ultra-violet (EUV) at wavelengths \NeVIIIstrong\ and \NeVIIIweak\ \citep[e.g.][]{ver94a}, which are redshifted into the COS spectral range at $z \gtrsim 0.5$. Moreover, the \NeVIII\ ionisation fraction in collisional ionisation equilibrium (CIE) is highest at temperatures $T \approx 5\times10^5 \K$ \citep[e.g.][]{sut93a}. 

To date, though, only a handful of intervening (as opposed to `proximate to the QSO')  \NeVIII\ absorbers have been identified in UV absorption spectra at redshifts $z < 1$ \citep{sav05a,nar09b,nar11a,tri11a,mei13a}. Interestingly, all these absorbers have similar strengths, with column densities $\NNeVIII \sim 10^{14} \psc$, and they all show a simple kinematic structure (at the given sensitivity and spectral resolution), thus suggesting that the observed \NeVIII\ absorption is  produced in gas under rather uniform physical conditions \citep[but see][]{tri11a}. In particular, the narrow range of measured temperatures ($\log_{10} (T / \K ) \approx 5.4 - 5.7$) indicates, first, that \NeVIII\ absorbers are very sensitive probes of the thermal state of the absorbing gas; and second, that \NeVIII\ absorbers are reliable tracers of gas at $T \approx 5 \times 10^{5} \K$ \citep[][but see \citealt{opp13b} who argue that \NeVIII\ absorbers may trace fossil AGN proximity zones]{sav05a}. Furthermore, their relatively high hydrogen column densities ($\NH \sim 10^{19} - 10^{20} \psc$) suggest that they contain a substantial amount of baryons \citep{sav05a,nar09b}. In view of these findings, \NeVIII\ absorbers indeed seem a promising tool in the search for warm-hot, diffuse gas and the `missing baryons'  in the low redshift Universe.

A clear link between the intervening \NeVIII\ absorbers and this peculiar gas phase has, however, not as yet been observationally established due to the current small number of detections, and support from the theoretical side is hence called for. This is the motivation for the present paper.

Here, we investigate in detail the physical conditions of gas producing \NeVIII\ absorption at low redshift following two rather independent and complementary approaches. First, we make use of a cosmological simulation that samples a representative volume of the universe at $z=0.5$ and which includes many of the physical processes believed to be important for the production and distribution of metals in the Universe. From this simulation we extract the distribution of densities and temperatures, the baryon content and the level of enrichment of a statistically significant sample of \NeVIII\ absorbers. Additionally, we use this sample to make predictions for two canonical absorption-line statistical quantities: the distribution of \NeVIII\ column densities and the number of \NeVIII\ absorbers of a given strength per unit redshift. Second, we develop an analytic model to predict standard observables (i.e.\ \NeVIII\ column density, \NeVIII\ central optical depth $\tau_0$, rest-frame equivalent width $W_r$) of \NeVIII\ absorption features as a function of the density and the temperature of the absorbing gas. The fundamental assumptions underlying our analytic model are that the gas phase producing \NeVIII\ absorption: 1) is exposed only to the extragalactic UV background; 2) is in ionisation equilibrium; 3) is embedded within a larger gaseous structure in (quasi) local hydrostatic equilibrium; 4) has an inhomogeneous metal distribution, with the metals filling only a fraction $c_L$ of the volume projected along the \los\ across the structure. We present and discuss these two approaches in Sections \ref{sec:sim} and \ref{sec:model}, respectively.

The combined results from our simulation and our analytic model --which we discuss in Section \ref{sec:results}--, allow us to determine the gas phases in the low redshift Universe that are expected to contain significant amounts of \NeVIII\  (in term of mass), and to identify these gas phases through their \NeVIII\ absorption. Our ultimate goal is to aid the interpretation of current observational data (discussed in Section \ref{sec:compare}), and to provide a guide for future observations aiming to detect \NeVIII\ absorbing gas using absorption spectroscopy.

We refer the impatient reader to Section \ref{sec:sum}, where we summarise the central results of our study. A detailed presentation of the status of current \NeVIII\ detections at low redshift and other results that might be of interest only to the more specialised reader are left for the appendix.

%--------------------------------------------------------------------------------------------------------------------------------------------------------------------------------
\section{Low-redshift \NeVIII\ absorbing gas: simulation and analytic model} \label{sec:predict}

%--------------------------------------------------------------------------------------------------------------------------------------------------------------------------------
% FIGURE: 
\begin{figure}
\resizebox{1.05\colwidth}{!}{\includegraphics{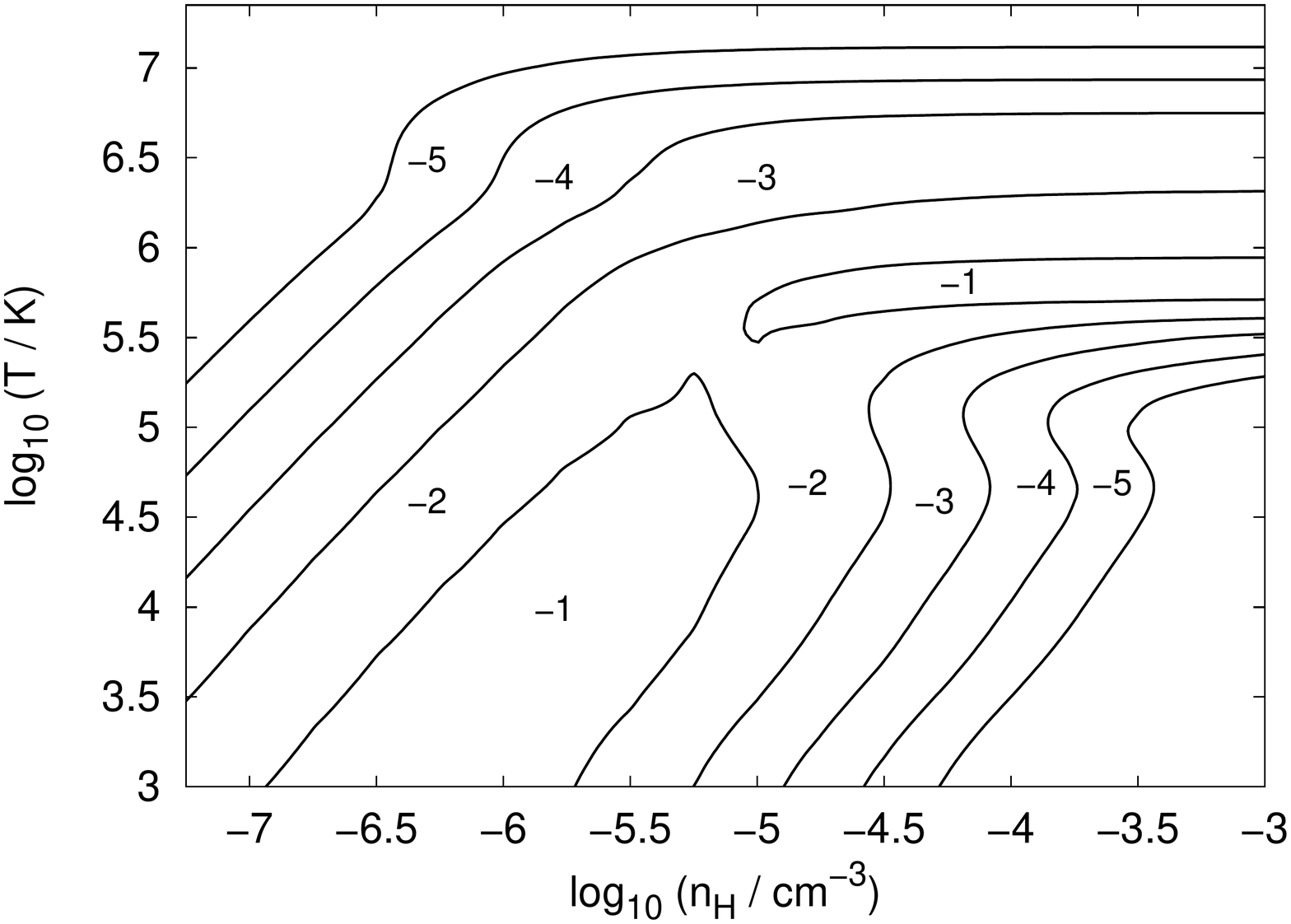}}
\caption{\NeVIII\ ionisation fraction, $\fNeVIII \equiv \fneviii$, of optically thin gas in ionisation equilibrium exposed to the \citet{haa01a} X-Ray/UV background radiation from galaxies and quasars at $z=0.5$. The numbers in the panel indicate the minimum $\fNeVIII $ (in logarithmic scale) in gas at the densities and temperatures within the corresponding contour. Note that relatively high ionisation fractions $\fNeVIII \sim 0.1$ occur in both low-density, low-temperature (photo-ionised) gas with $\nH < 10^{-5} \pcc$ and $T < 10^5 \K$, and in denser, warm-hot (collisionally ionised) gas with $\nH > 10^{-5} \pcc$ and $T \approx  5\times10^5 \K$.}
\label{fig:ion}
\end{figure}
%--------------------------------------------------------------------------------------------------------------------------------------------------------------------------------

Seven-times ionised neon (\NeVIII) can be produced by the ionisation of \NeVII\  \citep[ionisation potential $E_{p} = 207.271\pm0.013~{\rm eV}$; e.g.][]{kra06a} via the absorption of photons with energies $E_{\gamma} \geq E_{p}$ and by collisions with electrons and other atoms. The \NeVIII\ ion fraction $\fNeVIII \equiv \fneviii$ produced by both photo-ionisation and collisional ionisation in gas as a function of the gas density and temperature is presented in Figure \ref{fig:ion}, which we obtain from the tables provided by \citet{wie09a}. These tables, which were also used to calculate the ionisation balance in our simulation (see Section \ref{sec:sim}), were computed with the photoionisation package \textsc{cloudy} \citep[version 07.02.00 of the code last described by][]{fer98a} assuming the gas is exposed to the $z=0.5$ \citet{haa01a} model for the X-Ray/UV background radiation from quasars and galaxies (assuming a 10 per cent escape fraction for H-ionising photons). In addition, the gas is assumed to be dust-free, optically thin and in ionisation equilibrium. The most striking feature visible in Figure \ref{fig:ion} is the high \NeVIII\ ion fractions, $\fNeVIII \sim 0.1$, both in photo-ionised gas with $\nH < 10^{-5} \pcc$ and $T < 10^5 \K$, and in collisionally ionised gas with $\nH > 10^{-5} \pcc$ and $T \approx  5\times10^5 \K$. This is remarkable because it indicates that the presence of measurable \NeVIII\ absorption in spectra may by itself not be a sufficient condition for the detection of gas at $T > 10^5 \K$.

It is therefore important to explore in detail the connection between the physical conditions of the gas harbouring \NeVIII\ and the characteristics of the \NeVIII\ absorption it produces, which we do next using numerical simulations and analytic methods.

%--------------------------------------------------------------------------------------------------------------------------------------------------------------------------------
\subsection{Simulation} \label{sec:sim}

The simulation used in this work is one of the several cosmological simulations that together comprise the OverWhelmingly Large Simulations (OWLS) project, described in detail in \citet{sch10a}. These simulations were performed with a significantly extended version of the $N$-Body, Tree-PM, Smoothed Particle Hydrodynamics (SPH) code \textsc{gadget iii} -- which is a modified version of \textsc{gadget ii} \citep[last described in][]{spr05b} --, a Lagrangian code used to calculate gravitational and hydrodynamic forces on a system of particles.

Here we focus on a particular model from the OWLS suite referred to as {\em AGN}. This model adopts a flat $\Lambda$CDM cosmology characterised by the set of parameters $\{\Omega_{\rm m}, \, \Omega_{\rm b}, \, \Omega_{\Lambda}, \, \sigma_{8}, \, n_{\rm s}, \, h\} = \{ 0.238, \, 0.0418, \, 0.762, 0.74, \, 0.95, \, 0.73 \}$ as derived from the Wilkinson Microwave Anisotropy Probe (WMAP) 3-year data\footnote{These parameter values are largely consistent with the WMAP 7-year results \citep{jar11a}, the largest difference being the value of $\sigma_{\,8}$, which is $2 \, \sigma$ lower in the WMAP 3-year data than allowed by the WMAP 7-year data.} \citep{spe07a}. This model includes star formation following \citet{sch08e}, metal production and timed release of mass and heavy elements by intermediate mass stars, i.e., asymptotic giant-branch (AGB) stars and supernovae of Type Ia (SNIa), and by massive stars (core-collapse supernovae, SNIIe) as described by \citet{wie09b}. It further incorporates kinetic feedback by SNIIe based on the method of \citet{dal08b}, as well as thermal feedback by SNIa \citep{wie09b}. Radiative cooling by hydrogen, helium and heavy elements is included element by element following the method of \citet{wie09a}. The ionisation balance for each SPH particle is computed as a function of redshift, density, and temperature using pre-computed tables obtained with the photoionisation package \textsc{cloudy} \citep[version 07.02.00 of the code last described by][]{fer98a}, assuming the gas is exposed to the \citet{haa01a} model for the X-Ray/UV background radiation from galaxies and quasars. Additionally, our simulation includes feedback by active galactic nuclei (AGN) based on the model of black hole growth developed by \citet{boo09a}, which is a modified version of the model by \citet{spr05a}. Our simulation was run in a cubic box of $100 h^{-1}$ comoving mega-parsec (Mpc) on a side, containing $512^3$ dark matter (DM) particles and equally many baryonic particles. The initial particle mass is $4.1\times10^8 \hMsun$ (DM) and \mbox{$8.7\times10^7 \hMsun$} (baryonic). The gravitational softening is set to $8 ~h^{-1}$ comoving kilo-parsec (kpc) and is fixed at $2 h^{-1}$ proper kpc below \mbox{$z=3$}.

It is worth noting that this model has been shown to self-consistently reproduce several fundamental measurements, namely: the observed mass density in black holes at \mbox{$z=0$}; the black hole scaling relations \citep{boo09a} and their evolution \citep{boo11a}; the observed optical and X-ray properties, stellar-mass fractions, star-formation rates (SFRs), stellar-age distributions and the thermodynamic profiles of groups of galaxies \citep{mcc10a}; and the standard \HI\ observables (distribution of \HI\ column densities $\NHI$, distribution of Doppler parameters $\bhi$, \mbox{$\bhi - \NHI$} correlation) and the BLA line number density at low redshift \citep{tep12a}. For a more detailed description of this (and other) models that are part of the OWLS project, see \citet{sch10a}.

%--------------------------------------------------------------------------------------------------------------------------------------------------------------------------------
% FIGURE:
\begin{figure*}
{\resizebox{\textwidth}{!}{\includegraphics{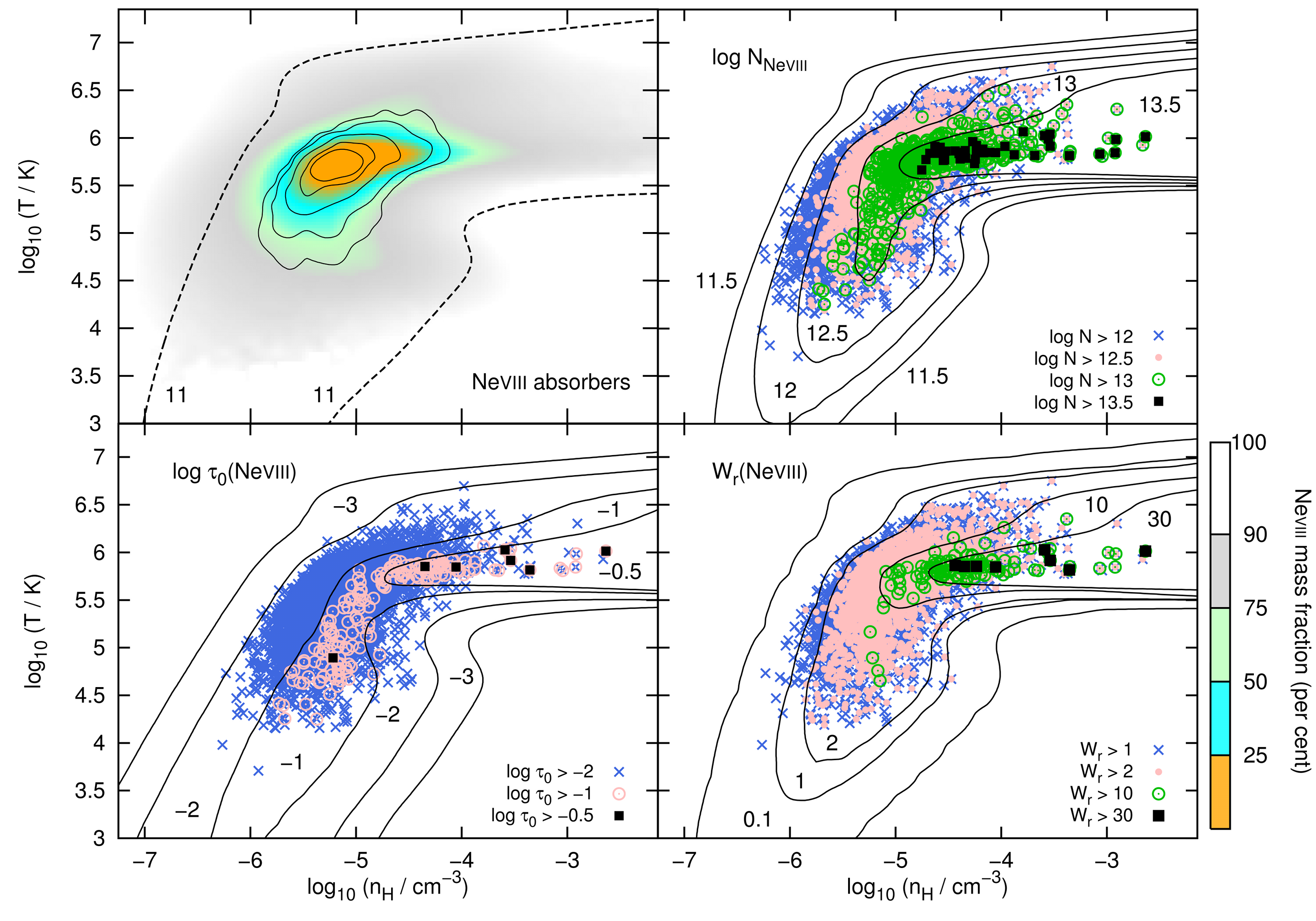}}}
\caption{{\em Top-left:} Distribution of \NeVIII\ mass in gas in our simulation at $z=0.5$. The coloured regions show the fraction of \NeVIII\ mass in gas contained within the corresponding region of $\nH - T$ space, as indicated by the colour bar in the bottom right margin. The solid contours show the distribution (by number) of gas densities and temperatures of the \NeVIII\ absorbers detected in a set of synthetic spectra at $z=0.5$; the outer-most contour contains 90 per cent of the absorbers. The dashed curves indicate the minimum column density $\log_{10} ( \NNeVIII / \psc ) = 11$ allowed in our fitting procedure. Note that virtually all the \NeVIII\ mass in gas and all the detected \NeVIII\ absorbers are contained within the region enclosed by this limit. {\em Top-right:} Scatter plot of \NeVIII\ absorbers symbol- and colour-coded by their \NeVIII\ column density. The solid contours indicate the \NeVIII\ column density (in $\psc$) predicted by our analytic model, equation \eqref{eq:nneviii_4}, adopting $[\Ne / \H ] = -0.7$ (relative to $\log_{10} \Xne_{\odot} = -3.91$), $f_g = 0.168$ and a filling fraction $\log_{10} c_L = -0.5$. {\em Bottom-left:} Scatter plot of \NeVIII\ absorbers symbol- and colour-coded by their central optical depth. The contours show the \NeVIII\ central optical depth predicted by our analytic model, equation (\ref{eq:tau0}; the  values for $[\Ne / \H ]$, $f_g$ and $c_L$ are identical to the values adopted before). {\em Bottom-right:} Scatter plot of \NeVIII\ absorbers symbol- and colour-coded by their rest-frame equivalent width (in m\AA). The contours indicate the \NeVIII\ rest-frame equivalent width in m\AA\ computed from the \NeVIII\ column density predicted by our analytic model, equation \eqref{eq:ew_1}. Note that the \NeVIII\ gas mass distribution in the top-left panel, and the contours in each panel are identical to the corresponding contours shown in Figure \ref{fig:predict}. The contours agree with the coloured points, indicating that our analytic model reproduces the simulation results very well.}
\label{fig:tune}
\end{figure*}
%--------------------------------------------------------------------------------------------------------------------------------------------------------------------------------

%--------------------------------------------------------------------------------------------------------------------------------------------------------------------------------
\subsubsection{Physical conditions of \NeVIII\ absorbing gas} \label{sec:phys}

We investigate the physical state of the gas containing \NeVIII\ in our simulation at $z=0.5$ in two different ways. Our first method is to compute the distribution of \NeVIII\ mass in gas as a function of density and temperature using the density, temperature, and \NeVIII\ mass of each SPH particle. For the second method we draw a number (5000) of random \loss\ through our simulation box and generate and fit a corresponding set of high-S/N, COS-resolution spectra, using the procedure described in Appendix \ref{sec:spec}. In this way we obtain a sample of roughly 8670 \NeVIII\ absorption components and we assign each a characteristic (i.e.\ \NeVIII\ optical depth weighted ) density ($\nH$) and a characteristic temperature ($T$). To this end, we follow the procedure described in \citet[][their Section 5.1]{tep11a}. Briefly, we use our synthetic spectra to compute the density (temperature) weighted by the \NeVIII\ optical depth in redshift space along the \los\ as in \citet{sch99a}. We then identify the pixel corresponding to the absorption line's centre, smooth (i.e.\ average) over $\pm2$ pixels around this pixel, and assign this last optical depth weighted average density (temperature) to the absorber. Note that our synthetic spectra have a constant pixel size $\delta v = 3 \kms$, which corresponds to a physical linear path $\delta l \approx 33~\kpc$ at $z=0.5$, assuming that the velocity spread is due solely to the Hubble expansion (note that our spectra are computed taking into account also peculiar velocities and thermal broadening). We thus smooth over a range $\Delta v = 12 \kms$ or $132 ~\kpc$ around the line centre.

The distributions of densities ($\nH$) and temperatures ($T$) of the \NeVIII\ gas in our simulation at $z=0.5$ obtained from these two different approaches are compared in the top-left panel of Figure \ref{fig:tune}  (note that the other panels will be discussed in Section \ref{sec:model}). The distribution of \NeVIII\ mass as a function of $\nH$ and $T$ obtained from the density, the temperature, and the \NeVIII\ mass of each SPH particle is shown by the coloured areas. Each of these is colour-coded by the fraction of \NeVIII\ mass in gas contained within the corresponding region of $\nH - T$ space, according to the colour bar displayed in the bottom right margin. So, for example, the innermost region (orange) contains 25 per cent of the \NeVIII\ mass in the simulation; the cyan and the orange area together enclose 50 per cent of the \NeVIII\ mass, and so on. The black contours show the distribution of \NeVIII\ optical depth weighted densities and temperatures of the \NeVIII\ absorbers detected in our simulated spectra. These contours enclose, starting from the innermost, 20, 40, 60, 80, and 90 per cent of the total number of absorbers. Note that the physical conditions of the \NeVIII\ absorbing gas inferred from the spectra (black contours) match the corresponding intrinsic properties of the \NeVIII\ harbouring gas in the simulation (coloured regions) very well. This proves that our choice of assigning physical quantities to the absorbers identified in synthetic spectra by using optical depth weighted quantities is a reliable method to unequivocally link the absorption of the gas to its physical state.

We find that most of the \NeVIII\ mass in our simulation resides in gas with densities with densities $10^{-6} \lesssim (\nH / \pcc) \lesssim 10^{-4}$  (or roughly 1 -- 150 times the cosmic mean at $z=0.5$) and temperatures $10^5 \lesssim (T / \K) \lesssim 10^6$. Hence, even though the \NeVIII\ ion fraction can be high both in photo- and collisionally ionised gas (see Figure \ref{fig:ion}), our simulation predicts that most of the \NeVIII\ is produced by collisional ionisation in low-density gas in ionisation equilibrium. This result disagrees in part with \citet{opp12a}, who find that most of the \NeVIII\ mass in their simulation (albeit at $z = 1$) is also in gas with densities $\nH \sim 10^{-5} \pcc$, but at significantly lower temperatures, only slightly above $T \sim 10^4 \K$. We note that we do find in our simulation \NeVIII\ absorbers with $T < 10^5 \K$ (see scatter plots in Figure \ref{fig:tune}); however, these are generally weak and their relative number is low.

%--------------------------------------------------------------------------------------------------------------------------------------------------------------------------------
% FIGURE:
\begin{figure*}
{\resizebox{0.33\textwidth}{!}{\includegraphics{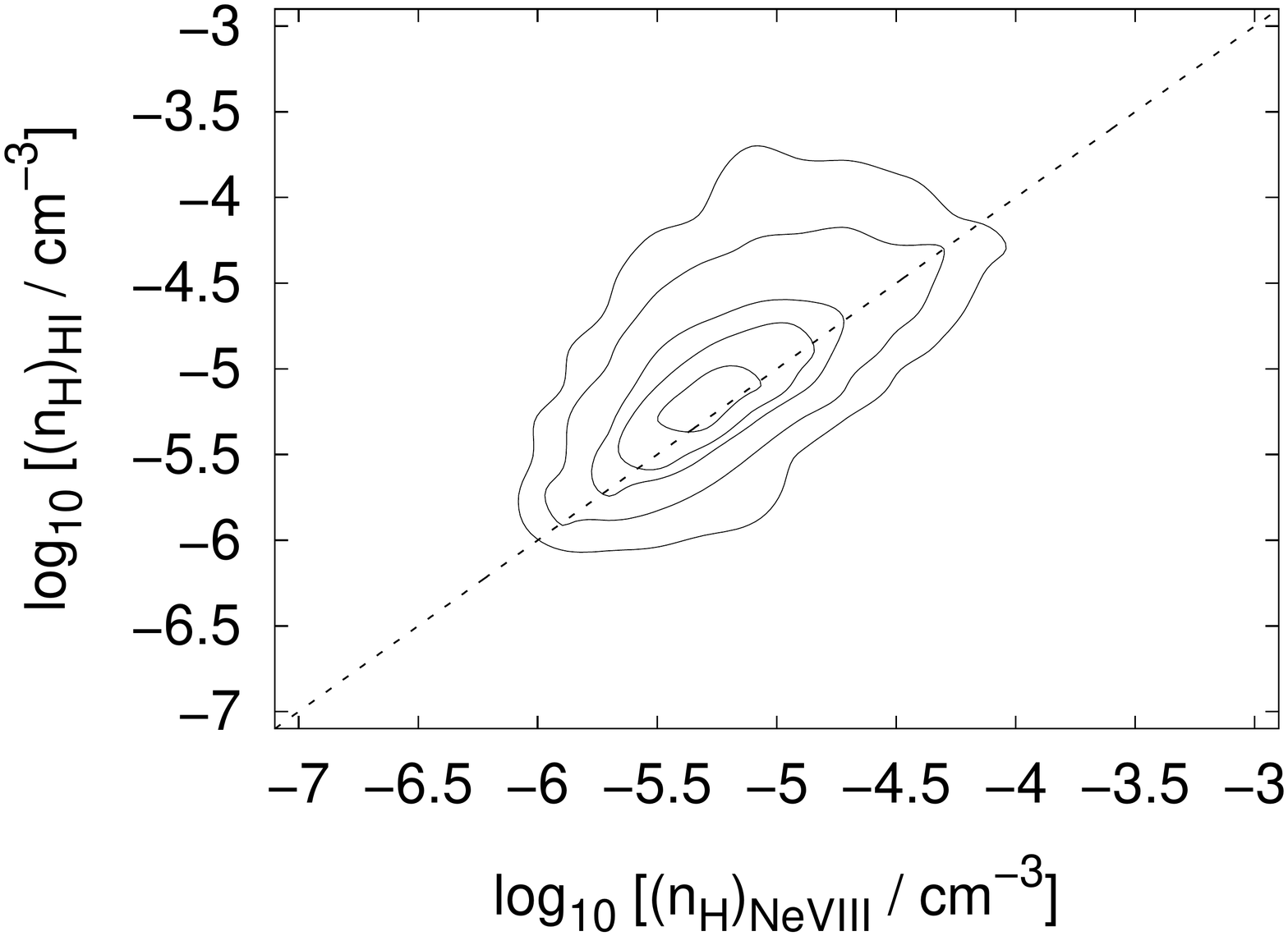}}}
{\resizebox{0.33\textwidth}{!}{\includegraphics{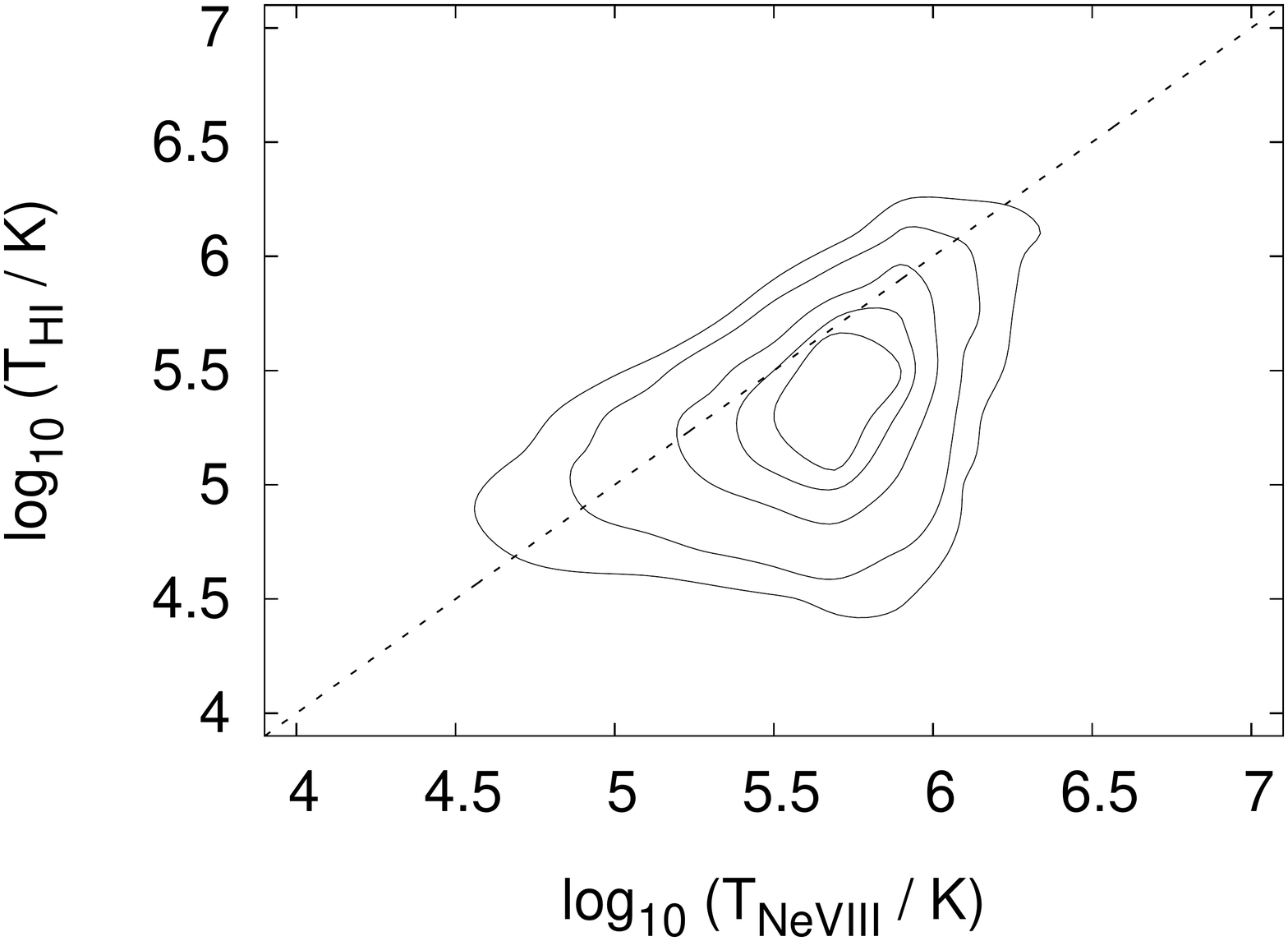}}}
{\resizebox{0.33\textwidth}{!}{\includegraphics{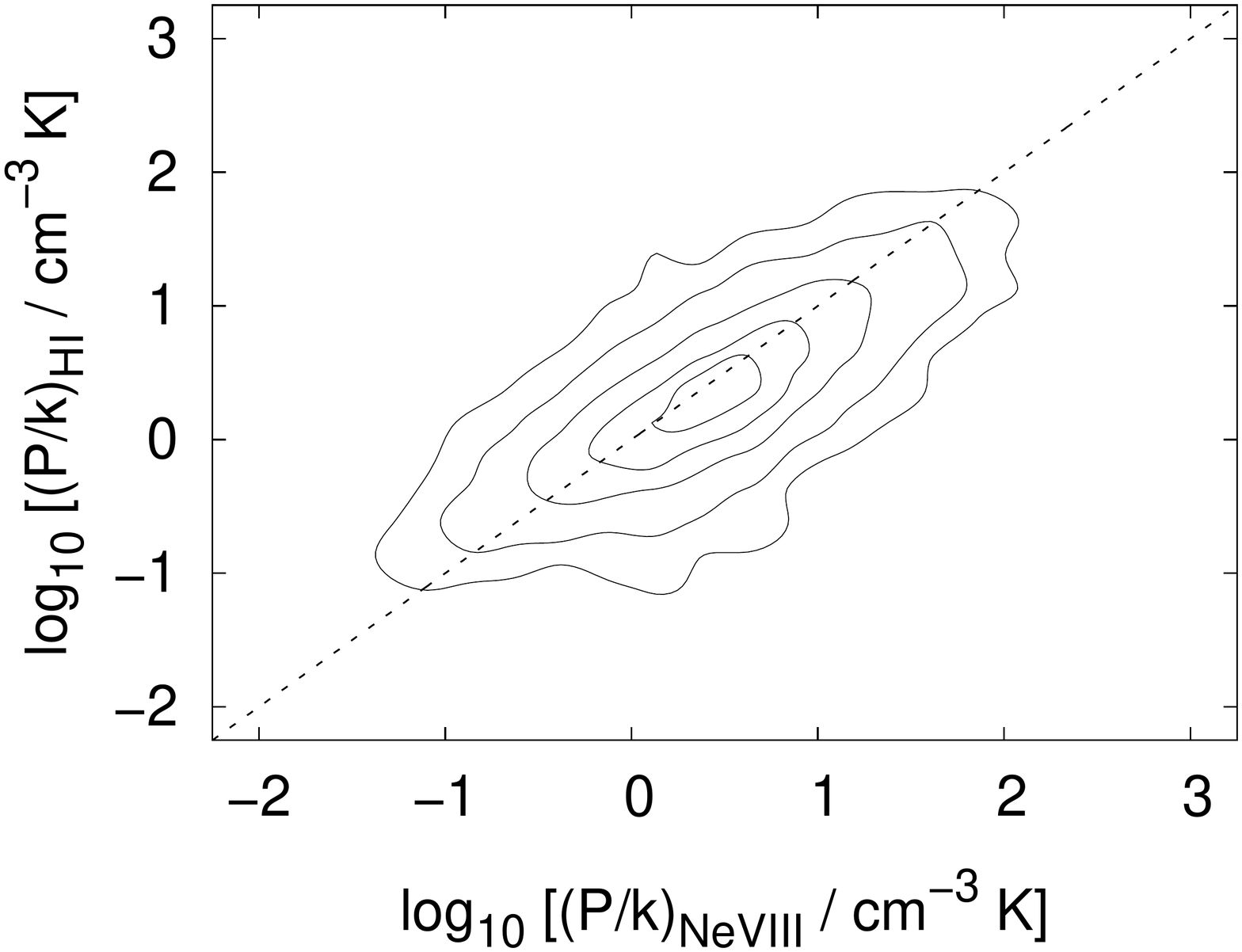}}}
\caption{Comparison of the density (left), temperature (centre), and thermal pressure (right) of the \NeVIII\ absorbing gas and that of the gas responsible for the \HI\ absorption at the same location in redshift space. The contours enclose, starting from the innermost, 20, 40, 60, 80, and 99 per cent of the absorbers (by number).}
\label{fig:phys}
\end{figure*}
%--------------------------------------------------------------------------------------------------------------------------------------------------------------------------------

A similar discrepancy concerning the typical temperature of \OVI\ absorbing gas between \citeauthor{opp12a}'s simulations and ours was reported in \citet{tep11a}. \citeauthor{opp12a} attribute the low fraction of highly-ionised metal absorbers with $T \gtrsim 10^5 \K$ in their simulation to the fact that their simulation does not resolve conductive interfaces within halos. Since our simulation does not resolve such interfaces either, the difference in the predicted temperature distribution of \NeVIII\ (or \OVI) absorbing gas most probably stems from the difference in the implementations of the sub-grid physics; in particular, the prescription to model winds on galactic scales may be crucial since these determine the distribution of mass and heavy elements among the different gas phases. Also, the assumption of ionisation equilibrium, which is also used by \citet{opp12a}, may be important \citep[e.g.][]{opp13a,opp13b} and future studies relaxing this assumption may obtain different results. We therefore advise that the physical conditions, in particular the thermal state, of highly ionised gas predicted by simulations should be taken with caution.

As will be explained in more detail in Section \ref{sec:model}, two of the implicit assumptions underlying our analytic model are that the density ($\nH$) and the temperature ($T$) of the \NeVIII\ absorbing clouds are comparable to the local density and the local temperature of the surrounding gas. This is reflected in the fact that the \NeVIII\ column density (and all the other derived spectral properties) as given by equation \eqref{eq:nneviii_4} depends on $\nH L_J$, where $L_J$ itself depends on $\nH$ and $T$  (see equation \ref{eq:jeans}). Moreover, since $\nH L_J \propto (\nH T)^{1/2} = (P/k)^{1/2}$ (where $P$ is the thermal pressure and $k$ is Boltzmann's constant), these assumptions imply that the metal absorbing gas is not far from thermal pressure equilibrium with its surrounding \HI\ absorbing medium.

In order to investigate if these conditions are satisfied by the \NeVIII\ absorbing gas in our simulation, we compare the characteristic density ($\nH$), temperature ($T$), and thermal pressure ($P/k = \nH T$) of each \NeVIII\ absorber in our sample with the corresponding quantity of the associated (i.e.\ aligned in velocity space) \HI\ absorber. To this end, we compute the \NeVIII- and the \HI-optical depth weighted density and temperature at the centre of each \NeVIII\ absorbing component identified in our simulated spectra as previously described. Note that when comparing \HI\ to \NeVIII\ weighted quantities, it is implicitly assumed that relative peculiar velocities between the \NeVIII\ and \HI\ absorbing phases are small, so that their alignment in velocity space implies their coincidence in physical space. The comparison of the densities, temperatures, and the thermal pressures thus obtained is shown in Figure \ref{fig:phys}. In each panel, the $x$-axis indicates the physical quantity (e.g.\ density) weighted by the \NeVIII\ optical depth and the $y$-axis indicates the corresponding quantity weighted by the \HI\ optical depth. The contours lines show the distribution (by number) of absorbers; these contours enclose, starting from the innermost, 20, 40, 60, 80, and 90 per cent of the total number of absorbers. The dotted line along the diagonal indicates in each case a perfect match between the two physical quantities considered.

The left panel of Figure \ref{fig:phys} shows that the density of the \NeVIII\ absorbing gas is similar to the density of the surrounding gas for most of the \NeVIII\ absorbers, being only slightly higher on average in the \HI\ absorbing phase with $\nH \gtrsim 10^{-5} \pcc$. In the middle panel we see that, while the temperature of both these phases is comparable for most of the absorbers ($\log_{10}(T / \K) \approx 5.5 - 5.7$), for a non-negligible fraction of the absorbers it is a factor $\sim10$ higher in the \NeVIII\ absorbing phase than in the \HI\ absorbing phase. This is expected since the \HI\ absorption, and thus the \HI\ optical-depth weighted temperature, will be dominated by the cooler absorbing \HI\ gas. The right panel demonstrates that the thermal pressure ($P/k$) in both phases is similar, and typically on the order $1 - 10  ~\pcc \K$. In other words, the deviations from pressure equilibrium between the gas responsible for the \NeVIII\ absorption and the associated \HI\ absorbing gas are not likely to be very large, at least for a significant fraction of the absorbers.

%--------------------------------------------------------------------------------------------------------------------------------------------------------------------------------
\subsubsection{Absorption line statistics} \label{sec:stats}

\begin{figure}
\resizebox{\colwidth}{!}{\includegraphics{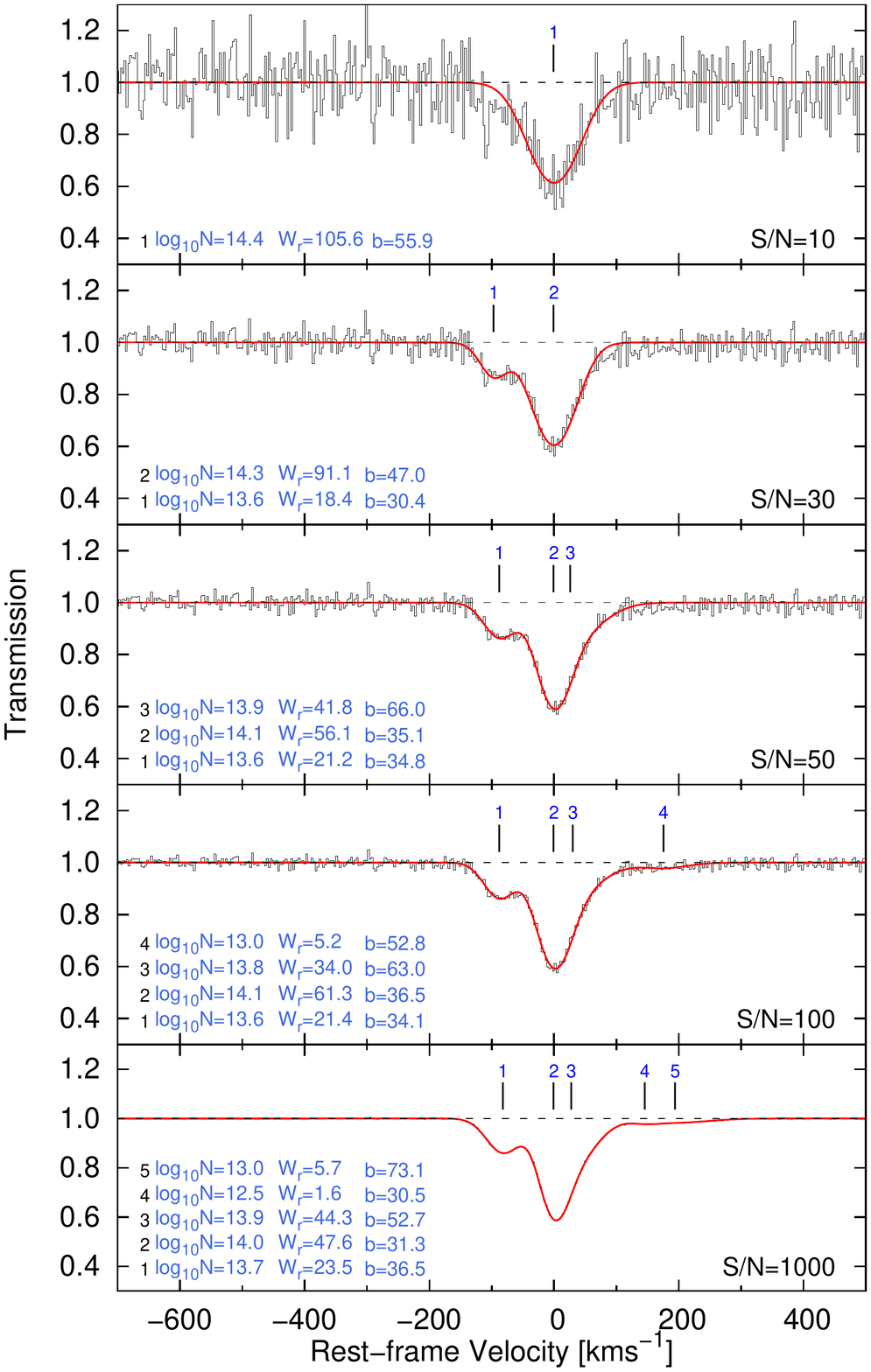}}
\caption{\NeVIII\ absorption feature with total \NeVIII\ column density $\log_{10} ( \NNeVIII / \psc) \approx 14.4$ along a random \los\ drawn from our simulation at $z=0.5$ in a synthetic COS-resolution spectrum adopting different S/N values (black histogram) and the corresponding fits (red). The vertical dashes mark the velocity centre of each fitted component; their corresponding logarithmic column density, rest-frame equivalent width (in m\AA), and Doppler parameter (in $\kms$) are listed in the bottom-left of each panel. While the equivalent width ($W_r$) and column density ($\NNeVIII$) integrated across the absorber are roughly the same for different S/N values ($W_r \approx 105~$m\AA\ and $\log_{10} ( \NNeVIII / \psc) \approx 14.4$, respectively), the values of $W_r$ and $\NNeVIII$ of the individual components vary substantially from panel to panel since a different number of components is required in each case to produce a good fit. Note that all fits have a reduced $\chi^2 \approx 1$ (see Appendix \ref{sec:spec} for details on the fitting procedure). This peculiar absorption feature is produced in gas with $\log_{10} ( \nH / \pcc) \approx -3.5$ and $\log_{10} (T / \K) \approx 6.1$.}
\label{fig:spec}
\end{figure}

\begin{figure}
\resizebox{\colwidth}{!}{\includegraphics{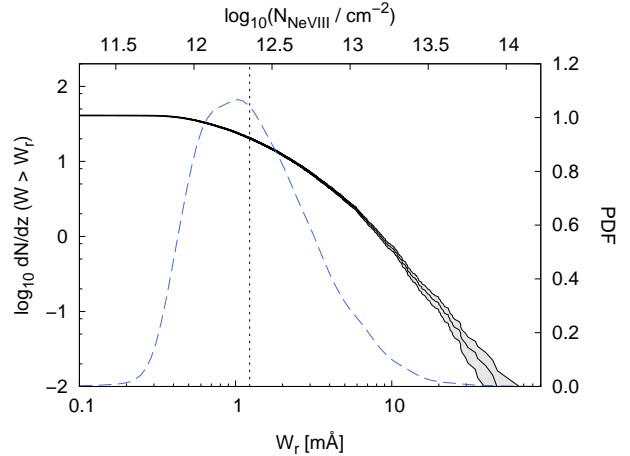}}
\caption{Number of \NeVIII\ absorption components per unit redshift above a given rest-frame equivalent width ($dN/dz(W > W_r)$; black curve) and the corresponding Poisson uncertainty (shaded area) obtained from a set of simulated COS-resolution spectra with S/N=1000 along 5000 random \loss\ drawn from our simulation at $z=0.5$. The top $x$-axis indicates the approximate \NeVIII\ column density corresponding to a given rest-frame equivalent width. The blue dashed curve shows the probability distribution of \NeVIII\ equivalent widths; its amplitude is indicated by the $y$-axis on the right. The vertical dotted line indicates the median value $W_r = 1.23~$m\AA. Note the small number density of absorbers with $W_r > 10~$m\AA.}
\label{fig:dndz}
\end{figure}

\begin{figure}
\resizebox{\colwidth}{!}{\includegraphics{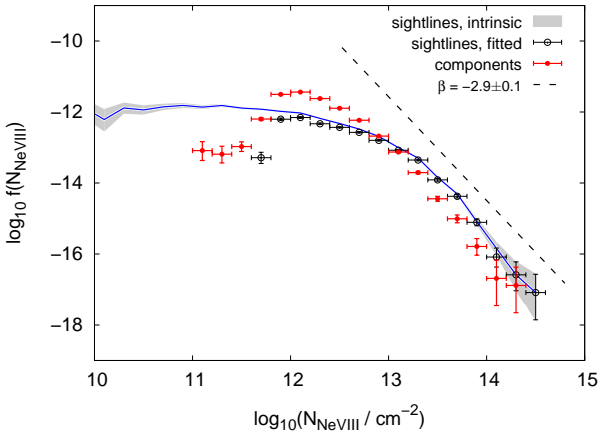}}
\caption{Distribution of \NeVIII\ column densities per unit $\NNeVIII$  per unit absorption distance, $f(\NNeVIII) \equiv \partial^2 \mathcal{N} / \partial \chi \partial \NNeVIII$ of the components identified in a set of simulated high-S/N, COS-resolution spectra along 5000 random \loss\ drawn from our simulation at $z=0.5$ (red symbols). The black symbols show the distribution of column densities integrated (i.e.\ adding up all the fitted components) along each spectrum; the blue curve show the distribution of {\em intrinsic} column densities integrated along each corresponding physical \los; the error bars along the $y$-axis and the shaded area indicate the corresponding Poisson scatter. The dashed line shows a fit to the red symbols in the range $\log_{10} (\NNeVIII / \psc) \in [12.5,~14.5]$ in the form of a power-law with slope $\beta = -2.9 \pm 0.1$. Note that the line has been shifted upwards for display purposes.}
\label{fig:cddf}
\end{figure}

Using the sample of \NeVIII\ absorption components obtained from our set of synthetic spectra (see Section \ref{sec:sim} and Appendix \ref{sec:spec}), we compute the \NeVIII\ line number density and the distribution of \NeVIII\ column densities at $z=0.5$ . The results are shown in Figures \ref{fig:dndz} and \ref{fig:cddf}, respectively, and discussed below. An example of a strong \NeVIII\ absorption feature produced in gas with $\log_{10} ( \nH / \pcc) \approx -3.5$ and $\log_{10} (T / \K) \approx 6.1$ along a random \los\ in our simulation at $z=0.5$ and its decomposition into individual Gaussian components is shown in Figure \ref{fig:spec} (see the figure caption for more details).

%--------------------------------------------------------------------------------------------------------------------------------------------------------------------------------
% FIGURE: 
%--------------------------------------------------------------------------------------------------------------------------------------------------------------------------------

%--------------------------------------------------------------------------------------------------------------------------------------------------------------------------------
% FIGURE: 
%--------------------------------------------------------------------------------------------------------------------------------------------------------------------------------

Figure \ref{fig:dndz} displays the number of \NeVIII\ absorption components per unit redshift with rest-frame equivalent width above a given threshold, $dN/dz(W > W_r)$, predicted by our simulation at $z=0.5$  (black) and the corresponding Poisson uncertainty (gray shaded area). The blue dashed curve shows the probability distribution (indicated by the right $y$-axis) of rest-frame equivalent widths, which has a median value $W_r = 1.23~$m\AA\ (indicated by the vertical dashed line). Thus, roughly half of the \NeVIII\ absorbers in our high-S/N (i.e.\ S/N=1000) simulated spectra have $W_r \lesssim 1~$m\AA, while the other half have a rest-frame equivalent width in the range $1 ~{\rm m\AA} < W_r  \lesssim 10 ~{\rm m\AA}$, with a very small fraction having $W_r \gtrsim 10~$m\AA, $dN/dz(W > 10 {\rm ~m\AA}) \sim 1$, and an even smaller fraction with $W_r > 30~$m\AA, $dN/dz(W > 30 {\rm ~m\AA}) \sim 10^{-1}$.

The value of the cumulative line frequency for absorbers with $W_r > 30~$m\AA\ we find is consistent with the predictions from other simulations \citep{opp12a}, but significantly lower than the values obtained from current observations. \citet{nar09b} give $dN/dz(W > 30 {\rm ~m\AA}) = 2.1$ (no error quoted), which they estimate indirectly from the \OVI\ line-number density at comparable redshifts, assuming every seventh \OVI\ absorber has associated \NeVIII\ absorption. Based on three detections along a single \los\ spanning $\Delta z = 0.43$, \citet{mei13a} obtain $dN/dz(W > 30 {\rm ~m\AA}) = 7^{+7}_{-4}$, which is only consistent with the value given by \citet{nar09b} at the lower $2\sigma$ level. For comparison, we estimate that the likelihood of finding three absorbers with $W_r \gtrsim 30 {\rm ~m\AA}$ over a redshift path $\Delta z =0.43$ in our simulation is approximately $1.5\times10^{-2}$.

The two values of $dN/dz$ inferred from observations rely on speculative assumptions and on very few detections along small total redshift paths all of which makes them highly uncertain. As noted by \citet{mei13a}, the three \NeVIII\ absorbers they detect are strongly clustered along the \los, which, as they suggest, could result from this \los\ piercing a region with a unusual high density of star-forming galaxies. Hence, the high number density of \NeVIII\ absorbers inferred from these studies might not be representative and likely overestimated as a result of selection effects and cosmic variance.  Increasing the sensitivity of the data would reveal weaker \NeVIII\ absorbers along many more \loss, thus increasing the number of detections but also the surveyed redshift path. It is, however, not clear whether the measured line frequency would converge to a value which is more in line with the result from our simulation. Alternatively, we consider the possibility that the low frequency of \NeVIII\ absorbers with $W_r \gtrsim 30 {\rm ~m\AA}$ (or $\log_{10} (\NNeVIII / \psc) \gtrsim 14$) in our simulation is an artefact related to the under-abundance of strong ($W_r \gtrsim 100$~m\AA)  \OVI\ absorbers in our \citep{tep11a} and other \citep[e.g.][]{opp09b} simulations that may result from turbulent mechanisms on sub-resolution scales which are not properly captured by these simulations. It is also possible that the observed strong \NeVIII\ (and \OVI) absorbers are not produced by gas in (collisional) ionisation equilibrium, but rather trace fossil AGN proximity zones which result from non-equilibrium effects \citep{opp13b}. Note that the assumption of ionisation equilibrium may break down even in the absence of fossil AGN radiation since the cooling time may be significantly shorter than the recombination time, at least for the strong absorbers. In any case, a significantly larger volume of data is required to get a reliable estimate of $dN/dz(W > W_r)$, which will allow a more robust comparison to the predicted value obtained from simulations like ours.\\

%--------------------------------------------------------------------------------------------------------------------------------------------------------------------------------
% FIGURE: 
%--------------------------------------------------------------------------------------------------------------------------------------------------------------------------------

The distribution of \NeVIII\ column densities per unit $\NNeVIII$  per unit absorption distance,\footnote{The absorption distance in a flat Universe dominated by matter and dark energy is given by
\begin{equation}
 d\chi =  (1+z)^2 [\Omega_m (1+z)^3 +\Omega_\Lambda]^{-1/2} dz \, . \notag
\end{equation}
} $f(\NNeVIII) \equiv \partial^2 \mathcal{N} / \partial \chi \partial \NNeVIII$, in our simulation at $z=0.5$ is shown in Figure \ref{fig:cddf}. The red symbols show the distribution of \NeVIII\ column densities (CDDF) of the components identified in a set of high-S/N, COS-resolution spectra along 5000 random \loss. The dashed line corresponds to the fit in the form of a power-law with slope $\beta = -2.9 \pm 0.1$ to the red symbols in the range $\log_{10} (\NeVIII / \psc) \in [12.5,~14.5]$, and it has been shifted upwards for display purposes. We note that the slope $\beta = -2.9 \pm 0.1$\ of the \NeVIII\ CDDF at $z=0.5$ is significantly steeper\footnote{The slope found by \citet{til12a} and \citet{tep11a} is obtained from the analysis of observed and simulated spectra, respectively, with moderate S/N: 10 -- 30. Using the set of high-S/N, COS-resolution synthetic spectra of this study, we find $\beta = -2.0 \pm 0.1$ for the \OVI\ CDDF, which is consistent with the slope obtained  in our previous work from simulated spectra with S/N = 10 -- 30.} than the slope $\beta = -2.1 \pm 0.1$ of the \OVI\ CDDF obtained from observations \citep{til12a} and found in our simulations at a somewhat lower redshift \citep[$z=0.25$; ][]{tep11a}.

The black symbols in Figure \ref{fig:cddf} show the distribution of column densities integrated (i.e.\ adding up the column densities of all identified components) along each spectrum of size $8340 \kms$ at $z=0.5$. The error bars along the $y$-axis indicate the corresponding Poisson uncertainty. The blue curve and the shaded area show the distribution of {\em intrinsic} \NeVIII\ column densities integrated along each corresponding physical \los\ of length $100 \hMpc$ (comoving) and its Poisson uncertainty, respectively. A perfect match between the black and the blue distributions would indicate that the totality of the \NeVIII\ (mass) is detected in absorption. The recovery is excellent for $\NNeVIII > 10^{12} \psc$. However, while the distribution of integrated, {\em intrinsic} column densities (blue / shaded) flattens off at column densities $\NNeVIII \sim 10^{12} \psc$, the distribution of column densities measured from the spectra (black / red) drops sharply\footnote{The fact that the \NeVIII\ detection limit  in our synthetic spectra with S/N=1000 and the flattening of the \NeVIII\ CDDF both correspond to $\NNeVIII \sim 10^{12} \psc$ is only coincidental and has therefore no deeper meaning.} at $\NNeVIII \lesssim 10^{12} \psc$. Thus, \NeVIII\ absorbers with $\NNeVIII \lesssim 10^{12} \psc$ should exist but their detection in absorption spectra requires S/N $>$ 1000. From the baryon content\footnote{The baryonic mass in the form of a particular ion is given by
\begin{eqnarray}
	\Omega_{ion} = \frac{m_{ion} }{\rho_{\rm c}} ~\left( \frac{c}{H_0} \Delta \chi_{tot} \right)^{-1} \N_{ion}^{\,tot}  && \nonumber \\
 	\approx 1.32\times10^{-10} \left( \frac{h}{0.73} \right)^{-1} A ~\Delta \chi_{tot}^{-1}  ~\left(\frac{\N_{ion}^{\,tot}}{10^{13} \psc} \right) \,, &&  \nonumber
\end{eqnarray}
where $A$ is the ion (atomic) mass number, $\Delta \chi_{tot} = \sum_{i=1}^{N_{\rm LOS}} \Delta \chi_{i}$ is the total surveyed absorption distance, and $\N_{ion}^{\,tot} = \sum_{i=1}^{N_{\rm LOS}} \sum_{j=1}^{N_{\rm abs}} \N_{ion}^{\,ij}$ is the total column density of all identified absorbers.
}
in the form of \NeVIII\ in the simulation, $\Omega_{\ionsubscript{Ne}{VIII}}^{sim} = 2.70\times10^{-8}$, and the baryon content in the form of \NeVIII\ recovered from the spectra, $\Omega_{\ionsubscript{Ne}{VIII}}^{fit}= 2.25\times10^{-8}$, we estimate that the fraction of \NeVIII\ mass in absorbers with $\NNeVIII < 10^{12} \psc$ is roughly 20 per cent. For comparison, \citet{mei13a} find $\Omega_{\ionsubscript{Ne}{VIII}} \approx  6\times10^{-8}$ at $z \approx 0.7$ along a single \los\ spanning a redshift path $\Delta \chi = 0.87$.

%--------------------------------------------------------------------------------------------------------------------------------------------------------------------------------
\subsection{Analytic model} \label{sec:model}

Now that we have discussed the physical conditions of the gas containing \NeVIII\ in our simulation, we proceed to investigate the connection between the physical state and the absorption signatures of this gas phase using an independent approach. As we have done in \citet{tep12a} for \HI, our goal here is to develop a model that allows us to predict the \NeVIII\ line observables (column density, equivalent width, central optical depth), and the spectral sensitivity in terms of S/N required to detect \NeVIII\ absorbing gas as a function of the gas density ($\nH$) and the gas temperature ($T$). Conversely, such a model should allow one to constrain the physical state of the absorbing gas from its observed absorption characteristics.\\

The \NeVIII\ central optical depth of an absorption line with column density $\NNeVIII$ and Doppler parameter $\bneviii$ is given by
\begin{eqnarray} \label{eq:tau0_neviii}
	\tau_{\,0}(\NeVIII) = \sqrt{\pi} ~r_{\,e} ~c ~\lambda_0 ~f_0 \left( \frac{\NNeVIII}{\bneviii} \right) & & \nonumber \\
	 \approx 0.596 ~\left(\frac{\NNeVIII}{10^{14} \cm^{-2}} \right) \left(\frac{\bneviii}{20 \kms} \right)^{-1} \, , & &
\end{eqnarray}
where $r_{\,e} = 2.82 \times 10^{-13}~\cm$ is the classical electron radius \citep{moh12a}, $c \approx 3.00\times10^{10} ~\cms$ is the speed of light in vacuum, and $\lambda_0$ and $f_0$ are the transition's rest wavelength and oscillator strength, respectively; the coefficient in the equation above is obtained for the \NeVIII\ $\lambda_0\!=$\NeVIIIstrong\ transition with $f_0 = 0.103$ \citep{ver94a}.

Since $\tau_0$, the equivalent width and the sensitivity in terms of S/N all depend on $\NNeVIII$ and $\bneviii$ (see equations \ref{eq:ew_1} and \ref{eq:sn} below), our problem reduces to finding expressions for these two quantities in terms of $\nH$ and $T$. As a starting point, we assume that the \NeVIII\ column density can be computed as
\begin{equation} \label{eq:nneviii_1}
	\NNeVIII = \int_{0}^{\, l} \nNeVIII ~dl' \, .
\end{equation}
Here, $l$ is the physical, linear extension of the gaseous, \NeVIII\ absorbing structure (or `cloudlet'), and $\nNeVIII$ is the corresponding \NeVIII\ particle density, which can be expressed in terms of the \NeVIII\ ion fraction, $\fNeVIII \equiv \fneviii$, and the neon abundance (by number) relative to hydrogen, $\Xne$, as
\begin{equation} \label{eq:nneviii_2}
	\nNeVIII = \left( \frac{\nNeVIII}{ \nNe} \right) \left( \frac{\nNe}{ \nH} \right) \nH = \fNeVIII ~\left( \frac{\nNe}{ \nH} \right) ~\nH \, .
\end{equation} 
Let us now assume\footnote{We have shown (\citealt{tep12a}; see also \citealt{alt11a,rah13c}) that our simulation (albeit at $z=0.25$) is consistent with the assumption that typical \HI\ absorbers are self-gravitating clouds in hydrostatic equilibrium with sizes $\sim L_J$.} that the structures responsible for most of the absorption are not far from local hydrostatic equilibrium and therefore typically have linear sizes on the order of the local Jeans length \citep{sch01a}
\begin{eqnarray} \label{eq:jeans}
	L_{\rm J} = 0.169 ~\Mpc ~( \nH / 10^{-5} \pcc )^{-1/2}  & & \nonumber \\
	\times ~(T / 10^4 \K )^{1/2} ~(f_g / 0.168 )^{1/2} \, , & & 
\end{eqnarray}
where, $f_g$ is the gas mass fraction and $f_g \equiv \Omega_b / \Omega_m = 0.168$ its universal value.\footnote{The total mass density parameter is \mbox{$\Omega_m = \Omega_b + \Omega_{\rm DM}$}. The most recent measurements of the  baryonic and dark matter density parameters yield, respectively, \mbox{$\Omega_b = 0.0449\pm0.0028$} and \mbox{$\Omega_{\rm DM} = 0.222 \pm 0.026$} \citep{jar11a}.}\\
We now need to take into account that the distribution of intergalactic metals is rather inhomogeneous\footnote{Note that the results by \citet{sch07a} correspond to observed metal-line absorbers at $z \approx 2.3$.} on scales $\lesssim 10^2 ~\kpc$ \citep{sch03a} and even on scales $\lesssim 1~\kpc$ \citep{sch07a}. We do this by introducing a factor, $c_L$, which gives the average filling fraction of the metal cloudlets along the gas cloud, in such a way that the total intersected metal absorbing path length (or the integrated linear size of the metal absorber) is
\begin{equation} \label{eq:size}
	l = c_L L_{\rm J} \, .
\end{equation}
Note that, by definition, $0 < c_L \leq 1$, where $c_L \ll 1$ implies a very inhomogeneous, and $c_L \approx 1$ a very homogeneous metal distribution across the gas structure.

If the physical conditions (i.e.\ temperature, density), the ionisation state and the neon abundance across the metal absorber are uniform, then equation \eqref{eq:nneviii_1} simplifies to 
\begin{equation} \label{eq:nneviii_3}
	\NNeVIII =  \fNeVIII ~\left( \frac{\nNe}{ \nH} \right)~c_L L_{\rm J} \nH \, .
\end{equation} 
Plugging equation \eqref{eq:jeans} into the above equation yields
\begin{eqnarray} \label{eq:nneviii_4}
	\NNeVIII =  5.90\times10^{14} \psc ~c_L ~10^{\, [\Ne/\H]} ~\left( \frac{\fNeVIII}{0.130} \right) & & \nonumber \\
	\times \left( \frac{\nH}{10^{-5} \pcc} \right)^{1/2} ~\left( \frac{T}{5 \times 10^5 \K}\right)^{1/2} ~\left( \frac{f_g}{0.168} \right)^{1/2} \, . & &
\end{eqnarray}
where \mbox{$[\Ne/\H] \equiv \log_{10}\Xne - \log_{10}\Xne_{\odot}$}. The fiducial values we have inserted in the above equation correspond to  the peak ionisation fraction of \NeVIII\ in CIE, $\fNeVIII = 0.130$ \citep{sut93a}, the (approximate) corresponding peak temperature, $T = 5\times10^5 \K$, and the solar neon abundance $\log_{10} \Xne_{\odot} = -3.91$ given by \citet[][their Table 2]{and89a}. Equation \eqref{eq:nneviii_4} readily shows that the \NeVIII\ column density of gas with $T \approx 5\times10^5 \K$ and $\nH \sim 10^{-5}$, and a neon abundance $[\Ne / \H ] = -0.5$ is $\NNeVIII \sim 10^{14} \psc$, assuming $f_g = 0.168$ and that the metals are homogeneously distributed across the structure, i.e.\ $c_L \equiv 1$.

In general, both thermal and non-thermal (e.g.\ turbulence, Hubble expansion) mechanisms will contribute to the total line width. If both types of mechanisms are stochastic and if they both obey a Gaussian distribution, then the total line width may be expressed in terms of the thermal ($b_T$) and non-thermal ($b_{nt}$) broadening as
\begin{equation} 
	\bneviii^2 = b_T^2 + b_{nt}^2 = \left(\frac{2 k T}{m_{\rm Ne}} \right) + b_{nt}^2 \, ,
\end{equation} 
where Boltzmann's constant $k = 1.38 \times 10^{-16} {\rm erg} \K^{-1}$, and $m_{\rm Ne} = 33.5 \times 10^{-24}~$g is the mass of the neon atom \citep{moh12a}. The line broadening due to non-thermal processes, in particular turbulent mechanisms, might be difficult to model. However, as we show in Appendix \ref{sec:spec} (see Figure \ref{fig:cog}), the intrinsic width of the \NeVIII\ absorption lines in our simulation at $z=0.5$ is dominated by thermal broadening, i.e.\ $b_T \gg b_{nt}$, such that the total line width may be well approximated by
\begin{equation} \label{eq:bneviii}
	\bneviii \approx b_T =  20.3 \kms \left( \frac{T}{5 \times 10^5 \K} \right)^{1/2} \, .
\end{equation} 
Using equations \eqref{eq:tau0_neviii}, \eqref{eq:nneviii_4}, and \eqref{eq:bneviii} we get
\begin{eqnarray} \label{eq:tau0}
	\tau_{\,0}(\NeVIII) = 3.46 ~c_L ~10^{\, [\Ne/\H]} ~\left( \frac{\fNeVIII}{0.130} \right) & & \nonumber \\
	\times \left( \frac{\nH}{10^{-5} \pcc} \right)^{1/2} \left( \frac{f_g}{0.168} \right)^{1/2} \, . & &
\end{eqnarray}
The above equation shows that, even for a density as low as $\nH \sim 10^{-5} \pcc$ and a metallicity $[\Ne / \H] = -0.5$, the opacity at the line centre of \NeVIII\ absorbing gas with a temperature close to the peak temperature of the \NeVIII\ ionisation fraction in CIE ($T \approx 5\times10^5 \K$) is high, $\tau_0 \sim 1$, provided that the metals are well mixed within the absorbing structure ($c_L \approx 1$).

Given a line strength $\tau_0$, we compute the corresponding rest-frame equivalent width using
\begin{equation} \label{eq:ew_1}
	W_r = \int_{-\infty}^{+\infty} \left(1 - {\rm e}^{-\tau(x)} \right) dx \, ,
\end{equation}
where $\tau(x) = \tau_0 \exp(-x^2)$, $x \equiv (\lambda - \lambda_0)/ \Delta \lambda_D$, and $\Delta \lambda_D = \lambda_0 (\bneviii / c)$.

Finally, the sensitivity in terms of the signal-to-noise ratio (S/N) required to detect gas at significance $N\sigma$ and the corresponding line strength are related by
\begin{equation} \label{eq:sn}
	N\sigma \sim N{\rm (S/N)}^{-1} \sim (1 - e^{-\tau_0}) \approx \tau_0\,, \qquad \tau_0 \ll 1 \, .
\end{equation}
Equations \eqref{eq:nneviii_4}, \eqref{eq:bneviii}, \eqref{eq:tau0}, \eqref{eq:ew_1}, and \eqref{eq:sn} thus constitute the solution to our problem. Of course, these equations are not restricted to \NeVIII; they can easily be modified to describe the absorption by any ion of any other metal.\\

In addition to the observables considered above, we provide an estimate of the total hydrogen column density ($\NH$) and the neutral hydrogen column density ($\NHI$) of the \NeVIII\ cloudlets using
\begin{equation} \label{eq:NH}
	\NH \equiv \nH ~l
\end{equation}
and
\begin{equation} \label{eq:NHI}
	\NHI \equiv \fHI ~\NH
\end{equation}
where $\fHI \equiv \fhi$ is the \HI\ ionisation fraction; this  is obtained from the same ionisation model we use to compute the \NeVIII\ ionisation fraction ($\fNeVIII$; see Section \ref{sec:predict}).

The actual observables, however, are the column densities of the gas {\em harbouring} the \NeVIII\ cloudlets, which we denote by $\NH' = \nH ~L_J$ and $\NHI' = \fHI ~\NH'$. Note that each of these quantities differs from the corresponding quantity (equations \ref{eq:NH} and \ref{eq:NHI}) by a factor $c_L^{-1}$, i.e.\ $\NH' = \NH ~c_L^{-1}$ and $\NHI' = \NHI ~c_L^{-1}$.\\

We close this section with an important remark: Strictly speaking, the particle density $\nH$ and the temperature $T$ {\em explicitly} entering equation \eqref{eq:nneviii_3} and equation \eqref{eq:bneviii}, respectively, correspond to the particle density and to the temperature of the metal  cloudlets, whereas the particle density and temperature {\em implicit} in equation (\ref{eq:nneviii_3}; and all derived equations) through $L_J \equiv L_J(\nH, T)$ refer to the particle density and to the temperature of the surrounding gas. These quantities need not be equal, in general. However, in deriving equations \eqref{eq:nneviii_4} and \eqref{eq:tau0} we have implicitly assumed they are (at least) of the same order of magnitude. As we have shown previously in Section \ref{sec:sim}, this assumption is well justified.\\

%--------------------------------------------------------------------------------------------------------------------------------------------------------------------------------
% FIGURE: 
\begin{figure*}
\resizebox{\colwidth}{!}{\includegraphics{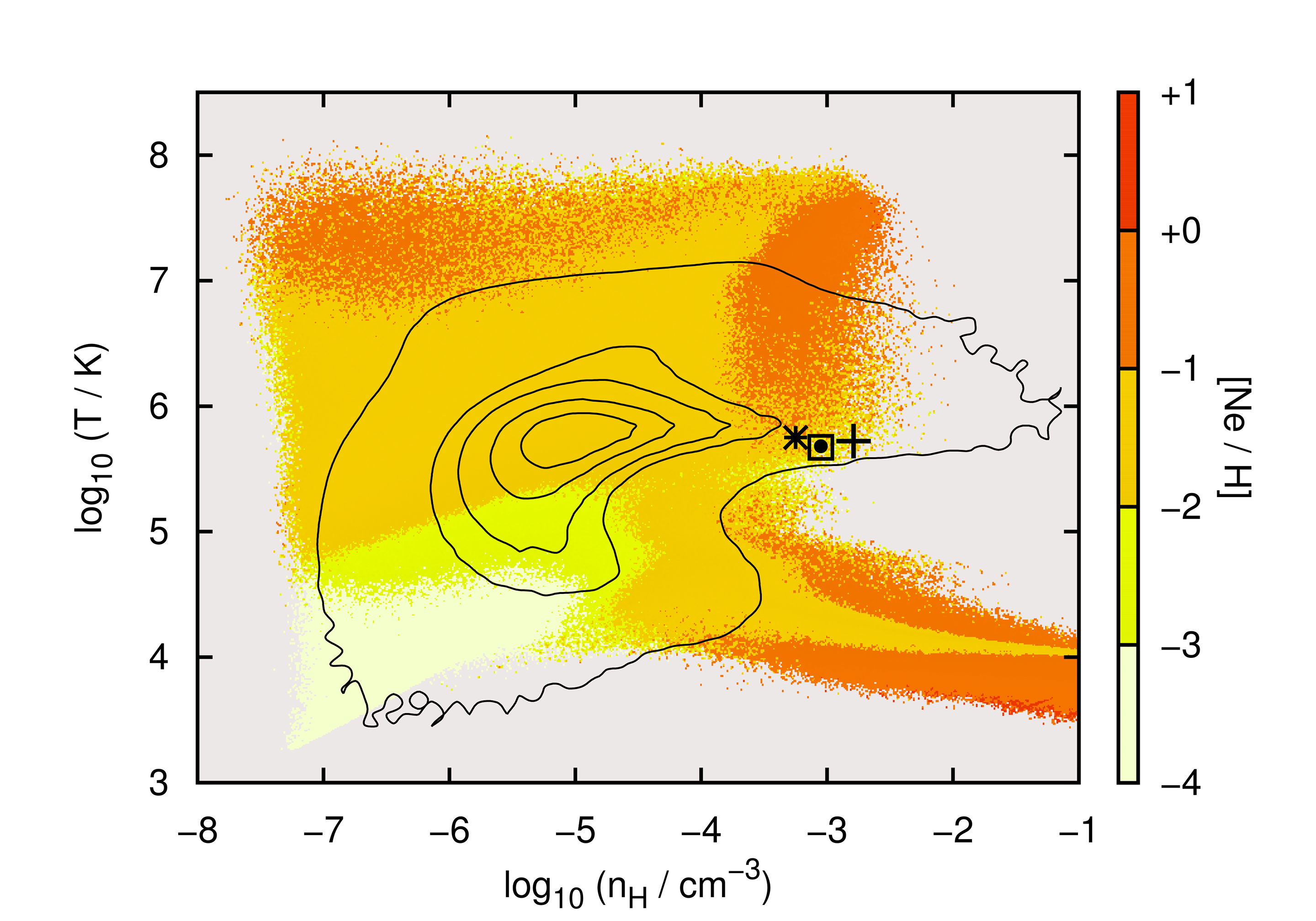}}
\resizebox{\colwidth}{!}{\includegraphics{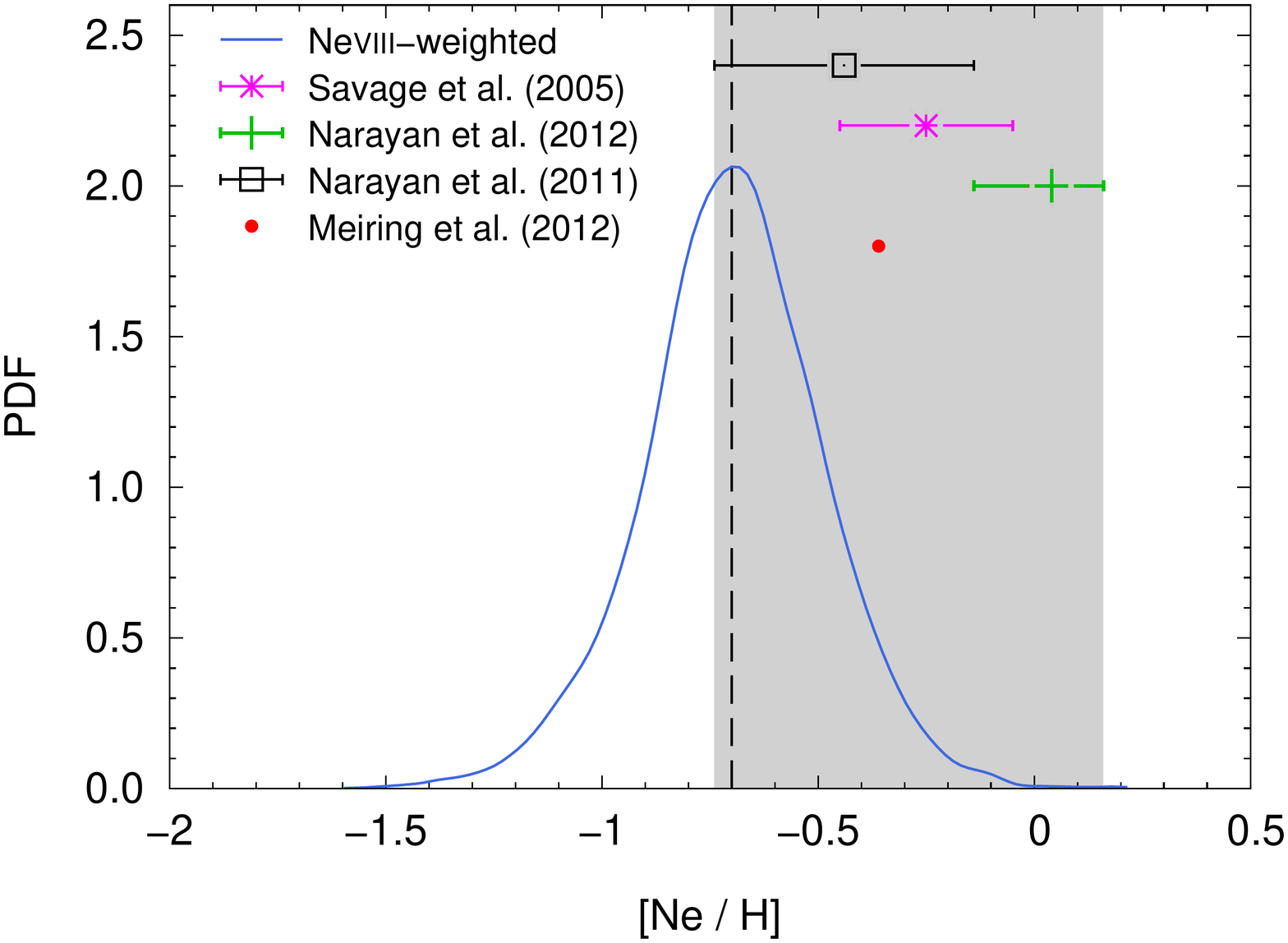}}
\caption{Neon abundance, $[\Ne/\H] \equiv \log_{10}\Xne - \log_{10}\Xne_{\odot}$, in gas in our simulation at $z=0.5$. {\em Left:} Mass-weighted mean of $[\Ne / \H]$ in gas as a function of the gas density and gas temperature. The black contours indicate the distribution of \NeVIII\ mass in gas, and they correspond to the distribution shown in colour in Figure \ref{fig:predict}. {\em Right:} Distribution of $[\Ne / \H]$ values in \NeVIII\ absorbing gas in our simulation (blue curve). The vertical dashed line indicates the corresponding median value $[\Ne / \H]$ = -0.7, which is also the fiducial value adopted in our analytic model. The shaded area indicates the full range of neon abundances in \NeVIII\ absorbers at low redshift inferred from observations (see Table \ref{tbl:data}). The data points in both panels represent a subset of the observed \NeVIII\ absorbers listed in Table \ref{tbl:data}. Note that their inferred neon abundances (right panel) are broadly consistent with the mean neon abundance of the gas in our simulation at the corresponding densities and temperatures (left panel). All abundances are given relative to $\log_{10} \Xne_{\odot} = -3.91$.}
\label{fig:meta}
\end{figure*}
%--------------------------------------------------------------------------------------------------------------------------------------------------------------------------------

\subsubsection{Setting the model parameters} \label{sec:param}

Our analytic model depends basically on two parameters:  1) $c_L$, which measures the homogeneity of the metal distribution within the gas structure; and 2) the (average) neon abundance $\Xne$ in the absorbing gas. The first parameter, $c_L$, is a free parameter that needs to be calibrated e.g.\ using simulations. The second parameter, $\Xne$, should be constrained by observations. Ideally, the value of $c_L$ obtained from our simulation should be consistent with observations as well; and likewise, the value of $\Xne$ constrained by observations should be consistent with the typical neon abundance of \NeVIII\ absorbers in our simulation.

Let us start by considering the second parameter. Figure \ref{fig:meta} shows the distribution of the neon abundance, $[\Ne/\H] \equiv \log_{10}\Xne - \log_{10}\Xne_{\odot}$, in gas in our simulation at $z=0.5$, adopting $\log_{10} \Xne_{\odot} = -3.91$. The left panel shows the mass-weighted average of the neon abundance in gas as a function of the gas density and temperature. The contours in this panel indicate the distribution of \NeVIII\ mass in gas, and correspond to the distribution shown in colour in the top-left panel of Figure \ref{fig:tune}.

The right panel displays the distribution of $[\Ne / \H]$ values in the \NeVIII\ absorbing gas in our simulation at $z=0.5$ (blue curve). The full range of neon abundances measured in some of the \NeVIII\ absorbers observed at low redshift is indicated by the shaded area. The data points in both panels represent a subset of the observed \NeVIII\ absorbers listed in Table \ref{tbl:data} (see Section \ref{sec:compare}). This figure shows that the neon abundance of the \NeVIII\ absorbers in our simulation is distributed over the range $-1.5 \lesssim [\Ne/\H] \lesssim 0$, with median $[\Ne / \H ] = -0.7$. Clearly, the average neon abundance of the gas in our simulation is broadly consistent with the neon abundance of the observed \NeVIII\ absorbers (data points) at similar densities and temperatures, although our simulation favours slightly lower values. In principle we could tabulate the mass-weighted mean of $\Xne$ obtained from our simulation as a function of $\nH$ and $T$ and use it as input in our analytic model. However, for simplicity and as a first approach, we have chosen to fix the neon abundance in our analytic model to the median value $[\Ne / \H ] = -0.7$ found in \NeVIII\ absorbing gas in our simulation. This choice is supported by the fact that the metallicity of the warm-hot gas in our simulations is robust to changes in the sub-grid physics and  evolves only slowly \citep{wie11a}.

Next we need to set the value of the parameter $c_L$. As mentioned before, this is a free parameter whose exact value cannot be set a priori. However, it is possible to constrain the range of values beforehand. For instance, by definition a value $c \approx 1$ implies that the metals are homogeneously distributed across the structure; also, by virtue of equations \eqref{eq:jeans}, \eqref{eq:size}, and \eqref{eq:nneviii_4}, such a value implies that the sizes of absorbers with $\NNeVIII \sim 10^{14}$ are on the order of 1~Mpc, neither of which is consistent with observations (see Section \ref{sec:compare}). Thus, values in the range $c_L \ll 1$ seem more plausible.

To determine the value of $c_L$ more precisely, we choose the following approach: First, we measure the \NeVIII\ column density ($\NNeVIII$), the rest-frame equivalent width ($W_r$) and the central optical depth ($\tau_0$) for each \NeVIII\ absorber identified in the synthetic spectra obtained from our simulation, as explained in Appendix \ref{sec:spec}. We use for this purpose the sample of absorbers identified in simulated spectra that have {\em not been convolved} with a COS line-spread function to avoid a potential bias between intrinsic/physical- and spectral properties (most importantly between the gas temperature and the corresponding line width; see Figure \ref{fig:cog}). Next, we extract subsets from our sample of \NeVIII\ absorbers chosen in terms of various $\NNeVIII$, $\tau_0$, and $W_r$ cuts, and compute for each of these \NeVIII\ absorbers a characteristic (i.e.\ \NeVIII\ optical depth weighted) density $\nH$ and a characteristic temperature $T$, as described in Section \ref{sec:sim}. We finally adjust $c_L$ in our analytic model (equations \ref{eq:nneviii_4}, \ref{eq:bneviii}, \ref{eq:tau0}, and \ref{eq:ew_1}) to simultaneously reproduce the distribution of each of these subsets on the $\nH - T$ plane.

We find that $\log_{10} c_L = -0.5$ (i.e.\ $c_L \approx 0.3$) yields very satisfactory results, adopting our fiducial values $[\Ne / \H ] = -0.7$ and $f_g = 0.168$, as shown in Figure \ref{fig:tune}. The panel on the top-right of this figure displays a scatter plot of the \NeVIII\ absorbers identified in our simulated spectra symbol- and colour-coded for the various $\NNeVIII$ cuts indicated in the legend. The solid contours indicate the predicted \NeVIII\ column density as a function of $\nH$ and $T$ as given by our analytic model (equation \ref{eq:nneviii_4}). The bottom-left and bottom-right panels display, respectively, a scatter plot of the \NeVIII\ absorbers symbol- and colour-coded by different \NeVIII\ central optical depth ($\tau_0$) and  rest-frame equivalent width ($W_r$) cuts. The contours in the corresponding panel show $\tau_0$ as a function of $\nH$ and $T$ as given by equation \eqref{eq:tau0} and $W_r$ as a function of $\nH$ and $T$ computed using equation \eqref{eq:ew_1}. The top-left panel displays the distribution of \NeVIII\ mass in gas as a function of its density and temperature in our simulation and has already been discussed in Section \ref{sec:sim}. The blue curve indicates the minimum column density $\NNeVIII = 10^{11} \psc$ allowed in our fitting procedure (see Appendix \ref{sec:spec}) and  has been computed using equation \eqref{eq:nneviii_4}. Note that virtually all the \NeVIII\ mass in gas is contained within the region on the $\nH - T$ plane defined by this limit. This is important as a consistency check because it implies that we can detect the bulk of the \NeVIII\ mass in absorption using our synthetic spectra.

The contours in each of the panels 2 - 4 (counting clockwise from the top-right) of Figure \ref{fig:tune} agree well with the values of the coloured points in the corresponding panel, indicating that our analytic model reproduces the spectral properties of a statistical significant sample of simulated \NeVIII\ absorbers as a function of the density and temperature of the absorbing gas. While it is true that our fiducial parameter values ($f_g = 0.168$, $\log_{10} c_L = -0.5$, $[\Ne / \H ] = -0.7$) may not apply in general to all absorbers, they should still be representative. First, there is no evidence to date that the ratio of baryonic to total mass in such systems should be very different from its universal value. Second, the value $\log_{10} c_L = -0.5$ yields sizes for the \NeVIII\ absorbers (see equation \ref{eq:size}) on the order of $10 - 10^2~\kpc$, which are consistent with the typical size of \NeVIII\ absorbers around halos both in observations \citep{mei13a} and in other simulations \citep{for13a}. Values $c_L \ll 1$ are also consistent with observations of \CIV\ absorbers at high redshift \citep{sch07a}. And finally, the fiducial value of the neon abundance is broadly consistent with the currently measured abundances of \NeVIII\ absorbers at low redshift.

In principle, higher values of $[\Ne / \H]$ and lower values of $c_L$ are both allowed by observations. Within the framework of our model, however, these parameters are not completely independent of each other. For instance, if we demand that our analytic model yields quantitatively the same results in order to remain consistent with our simulation, then a higher neon abundance requires a lower value of $c_L$ for a given $\nH$ and $T$. In other words, our analytic model predicts that at a given \NeVIII\ column density, high (low) metallicity absorbers are more compact (extended), on average.

Needless to say, it is not the {\em exact} value of $c_L$ (nor of $[\Ne / \H ]$), but its order of magnitude that is important as a parameter that quantifies the degree of metal mixing in gas. The power of our analytic model ultimately lies in providing a connection between the observable parameters ($\NNeVIII$, $W_r$, $\NHI$) and the physical parameters ($\nH$, $T$, $\NH$, $[{\rm X} / \H]$, $l$) of the absorbing gas, while accounting for the observed inhomogeneous distribution of metals.

%--------------------------------------------------------------------------------------------------------------------------------------------------------------------------------
\section{Spectral signatures of \NeVIII\ absorbing gas} \label{sec:results}

\begin{figure*}
{\resizebox{\textwidth}{!}{\includegraphics{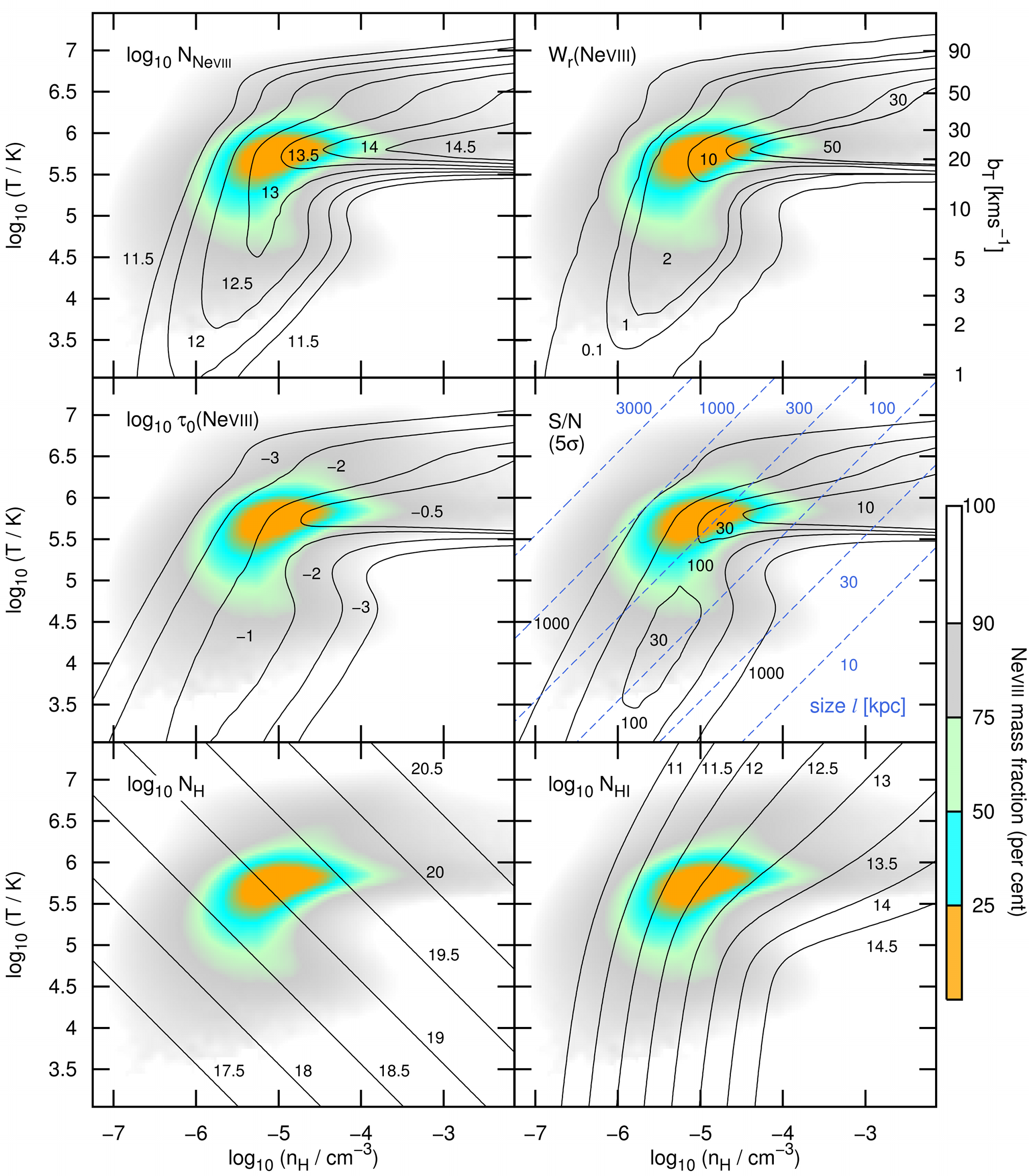}}}
\caption{\NeVIII\ line observables and mass distribution of \NeVIII\ in gas at $z=0.5$ as a function of hydrogen particle density ($\nH$) and temperature ($T$). The coloured regions show the fraction of \NeVIII\ mass in the simulation indicated by the colour bar on the bottom-right margin, and they are identical in all panels (and identical to the coloured regions shown in Figure \ref{fig:tune}). The contours in each panel show the predictions of our analytic model. {\em Top-left:} \NeVIII\ column density ($\NNeVIII$ in $\psc$) as given by equation \eqref{eq:nneviii_4}. {\em Top-right:} Rest-frame equivalent width (in m\AA) of a \NeVIII\ absorption line computed using equation \eqref{eq:ew_1}. The right $y$-axis in this panel indicates the corresponding thermal broadening (equation \ref{eq:bneviii}). {\em Middle-left:} \NeVIII\ central optical depth ($\tau_0$) as given by equation (\ref{eq:tau0}). {\em Middle-right:} Minimum S/N required to detect \NeVIII\ absorbing gas calculated assuming a $5\sigma$ significance, assuming the line to be fully resolved. The dashed contours indicate the approximate size (in kpc) across the \NeVIII\ absorber. {\em Bottom-left:} Total hydrogen column density ($\NH$ in $\psc$) of  \NeVIII\ bearing gas (equation \ref{eq:NH}). {\em Bottom-right:} Neutral hydrogen column density ($\NHI$ in $\psc$) of \NeVIII\ bearing gas  (equation \ref{eq:NHI}). Note that the linear size, and the total H and \HI\ column densities of the gas clouds that harbour the metal patches are a factor $c_L^{-1}$ greater. The contours shown in all panels have been computed assuming a neon abundance $[\Ne / \H ] = -0.7$ (relative to $\log_{10} \Xne_{\odot} = -3.91$), a gas fraction $f_g = 0.168$, and a line of sight metal filling fraction $\log_{10} c_L = -0.5$ (i.e.\ $c_L \approx 0.3$).}
\label{fig:predict}
\end{figure*}

Now that we have found sensible values for the parameters entering our analytic model, we use it, in combination with our simulation, to analyse the connection between the spectral characteristics and the physical state of \NeVIII\ absorbing gas at $z=0.5$. The results are shown in Figure \ref{fig:predict} and discussed below.

Each of the panels in this figure displays a series of contours that show the behaviour predicted by our analytic model for \NeVIII\ absorbers as a function of the density and temperature of the \NeVIII\ absorbing gas. The spectral and physical parameters we consider are: the \NeVIII\ column density (top-left), the \NeVIII\ rest-frame equivalent width (top-right), the \NeVIII\ central optical depth (middle-left), the minimum S/N (at $5\sigma$ significance) required to detect \NeVIII\ absorbing gas assuming the line to be fully resolved (middle-right), the total hydrogen column density $\NH$ (bottom-left), and the neutral hydrogen column density $\NHI$ (bottom-right) of the \NeVIII\ bearing gas. The alternative $y$-axis in the top-right panel indicates the thermal broadening of a \NeVIII\ absorption line (equation \ref{eq:bneviii}). The dashed lines in the middle-right panel indicate the approximate size (in kpc) across the \NeVIII\ absorber, given by equation \eqref{eq:size}. The column density, the central optical depth, and the rest-frame equivalent width are computed using equations \eqref{eq:nneviii_4}, \eqref{eq:tau0}, and \eqref{eq:ew_1}, respectively, adopting $\log_{10} c_L = -0.5$ (i.e.\ $c_L \approx 0.3$) and the fiducial values $[\Ne / \H ] = -0.7$ and $f_g  = 0.168$. The coloured areas (which are identical in all panels) show the distribution of \NeVIII\ mass in gas in our simulation at $z=0.5$ and have already been discussed in Section \ref{sec:sim}. Note that they are identical to the distribution shown in colour in the top-left panel of Figure \ref{fig:tune}.

%--------------------------------------------------------------------------------------------------------------------------------------------------------------------------------
% FIGURE:
%--------------------------------------------------------------------------------------------------------------------------------------------------------------------------------

Quite apparent is the similarity between the behaviour of the \NeVIII\ ion fraction ($\fNeVIII$; Figure \ref{fig:ion}) and the behaviour of each of the line observables with $\nH$ and $T$, indicating that $\fNeVIII$ is the dominant factor in setting the absorption signatures of the absorbing gas. Yet, the correspondence is not perfect. In particular, we see that despite the relatively high ion fraction $\fNeVIII \sim 0.1$ of low-density, photo-ionised gas with $\nH \sim 10^{-6} \pcc$ and $T \sim 10^4 \K$, the corresponding \NeVIII\ column density is low, $\NNeVIII \sim 10^{12} \psc$, compared to the \NeVIII\ column density $\NNeVIII \sim 10^{14} \psc$ of gas with $\nH \sim 10^{-4} \pcc$ and $T > 10^5 \K$. This can be understood in terms of equation \eqref{eq:nneviii_4}, which predicts that, for a fixed $\fNeVIII$ and a fixed neon abundance, the \NeVIII\ column density scales as $\NNeVIII \propto \nH~l \propto (\nH ~T)^{1/2}$. As a result, the \NeVIII\ column density of photo-ionised gas at $\nH < 10^{-5} \pcc$ and $T <10^5 \K$ can be orders of magnitude lower than the \NeVIII\ column density of collisionally ionised gas with $\nH > 10^{-5} \pcc$ and $T \approx 5\times10^5 \K$, even though $\fNeVIII \sim 0.1$ in both gas phases and even if we assume identical metallicities. The equivalent width ($W_r$; top-right panel) of the \NeVIII\ absorbing features produced by these gas phases behaves in a similar way, since for $\NNeVIII \lesssim 10^{14} \psc$, which is the case of much of the gas in the range of densities and temperatures shown, $W_r$ is directly proportional to $\NNeVIII$ (see right panel of Figure \ref{fig:cog}).

The optical depth (middle-left panel) of the gas shows a slightly different behaviour. For instance, the optical depth at the line centre of \NeVIII\ absorbers produced in photo-ionised and collisionally ionised gas is comparable, $\tau_0 \gtrsim 0.1$, even though the corresponding column densities are an order of magnitude apart. This can be explained as follows: At temperatures $T < 10^5 \K$ the \NeVIII\ absorption lines are expected to be narrow ($\bneviii < 10 \kms$; see right $y$-axis in top-right panel) if their width is dominated by thermal broadening. Since $\tau_0 \propto (\NNeVIII / \bneviii)$, a decrease in $\NNeVIII$ is counterbalanced by a decrease in $\bneviii$ such that $\tau_0$ remains approximately constant. Since the minimum S/N value required to detect \NeVIII\ in gas at a given density and temperature scales as $(\tau_0)^{-1}$ (see equation \ref{eq:sn}), there is a natural correspondence between low (high) optical depths of the \NeVIII\ absorbing gas and the high (low) S/N values required to detect such gas, as can be seen by comparing the middle panels with each other.

Finally, the total hydrogen column density $\NH$ scales as $(\nH T)^{1/2}$ (see equation \ref{eq:NH}), which explains the linear (increasing) behaviour of $\log_{10} \NH$ with $\log_{10} \nH$ ($\log_{10} T$) for a fixed $T$ ($\nH$) as seen in the bottom-left panel. Likewise, the typical linear size across the \NeVIII\ absorber shown by the blue dashed lines in the middle-right panel obeys $l \propto L_J \propto (T / \nH)^{1/2}$, thus linearly increasing (decreasing) with $\log_{10} T$ ($\log_{10} \nH$). The behaviour of $\NHI$ is dictated by the non-trivial dependence of $\fHI$ on $\nH$, and particularly on $T$ (see equation A7 of \citealt{sch01a}, and Figure 7 of \citealt{tep12a}).

%--------------------------------------------------------------------------------------------------------------------------------------------------------------------------------
% Table: 
\begin{table*} 
\begin{center}
\caption{Summary of \NeVIII\ detections at $z < 1$. See text for details on the notes. Reference key: (1) \citet{sav05a,sav11b}; (2) \citet{nar09b,nar12a}; (3)  \citet{nar11a}; (4) \citet{mei13a}; (5) \citet{tri11a}.}
\label{tbl:data}
\begin{tabular}{lllllllrllc}
\hline
 QSO	& $z_{\rm abs}$	& $\log_{10} (\NNeVIII)$$^{a}$	& $\bneviii$$^{a}$	& $W_r$$^{c}$	& $\log_{10} \nH$$^e$	& $\log_{10} ( T )$		& $[\Ne / \H]$$^f$ 	& $\log_{10} ( \NH )$		&  $\log_{10} ( \NHI )$	& Reference	\\
		&				& $[\psc]$					& $[{\rm kms^{-1}}]$	& [m\AA]		& $[\pcc]$				& $[{\rm K}]$			&  				& $[\psc]$				& $[\psc]$				&			\\
\hline
 HE 0226-4110		&  0.207 & $13.89 \pm 0.11$	& $22.6 \pm 15.0$	& $36 \pm 10$		& -4.59		& 5.68		& -0.25	& 20.06				& 13.87		& (1) \\ 
 3C 263			& 0.326 & $14.06 \pm 0.08$	& $52.4 \pm 8.8$	& $57 \pm 10$		& -3.89		& 5.72		& 0.04	& 19.48				&  13.09		& (2) \\ 
 PKS 0405-123		& 0.495 & $13.96 \pm 0.06$$^{b}$	& --			& $45 \pm 6$$^{b}$	& -3.10$^d$	& 5.67$^d$	& -0.44	&19.70$^d$ 			& 13.50		& (3) \\ 
 PG 1148+549			& 0.684 & $13.95 \pm 0.04$	& $32.0 \pm 5.0$	& $43 \pm 4$		& -3.54	&  5.69		& -0.36	& 19.80				& 13.60		& (4) \\ 
 PG 1148+549			& 0.705 & $13.82 \pm 0.06$	& $28.2 \pm 7.1$	& $33 \pm 4$		& -3.75	&  5.69		& 0.14	& 19.00				& 12.80		& (4) \\ 
 PG 1148+549			& 0.725 & $13.81 \pm 0.06$	& $41.4 \pm 7.5$	& $33 \pm 4$		& -3.36	& 5.72		& 0.14	& 18.90				& 12.60		& (4) \\ 
 PG 1206+459		& 0.927 & $13.71 \pm 0.29$		& --				& --				& --			& 5.56		& --	& 19.80				& --			& (5) \\ 
 PG 1206+459		& 0.927 & $14.04 \pm 0.08$		& --				& --				& --			& 5.60		& --	& 20.30				& --			& (5) \\ 
 PG 1206+459		& 0.927 & $14.53 \pm 0.04$$^g$	& --				& --				& --			& 5.65		& --	& 20.50				& --			& (5) \\ 
\hline
\end{tabular}
\end{center}
\begin{list}{}{}
	\item[$^a$] Measured using Gaussian line profile fitting, unless stated otherwise.
	\item[$^b$] Measured using the apparent optical depth method \citep[AOD; ][]{sav91a}.
	\item[$^c$] Computed here using the measured \NeVIII\ column density and Doppler parameter assuming a Gaussian line profile, unless stated otherwise.
	\item[$^d$] Estimated assuming photo- and collisional ionisation equilibrium.
	\item[$^e$] Hydrogen density of the associated (i.e., aligned in velocity space) photo-ionised phase traced by low ions, e.g., \CIII, \SiIII.
	\item[$^f$] Relative to $\log_{10} \Xne_{\odot} = -3.91$.
	\item[$^g$] Total column density of a two-component feature.
\end{list}
\end{table*}
%--------------------------------------------------------------------------------------------------------------------------------------------------------------------------------

%--------------------------------------------------------------------------------------------------------------------------------------------------------------------------------
% FIGURE:
\begin{figure*}
{\resizebox{1.1\textwidth}{!}{\includegraphics{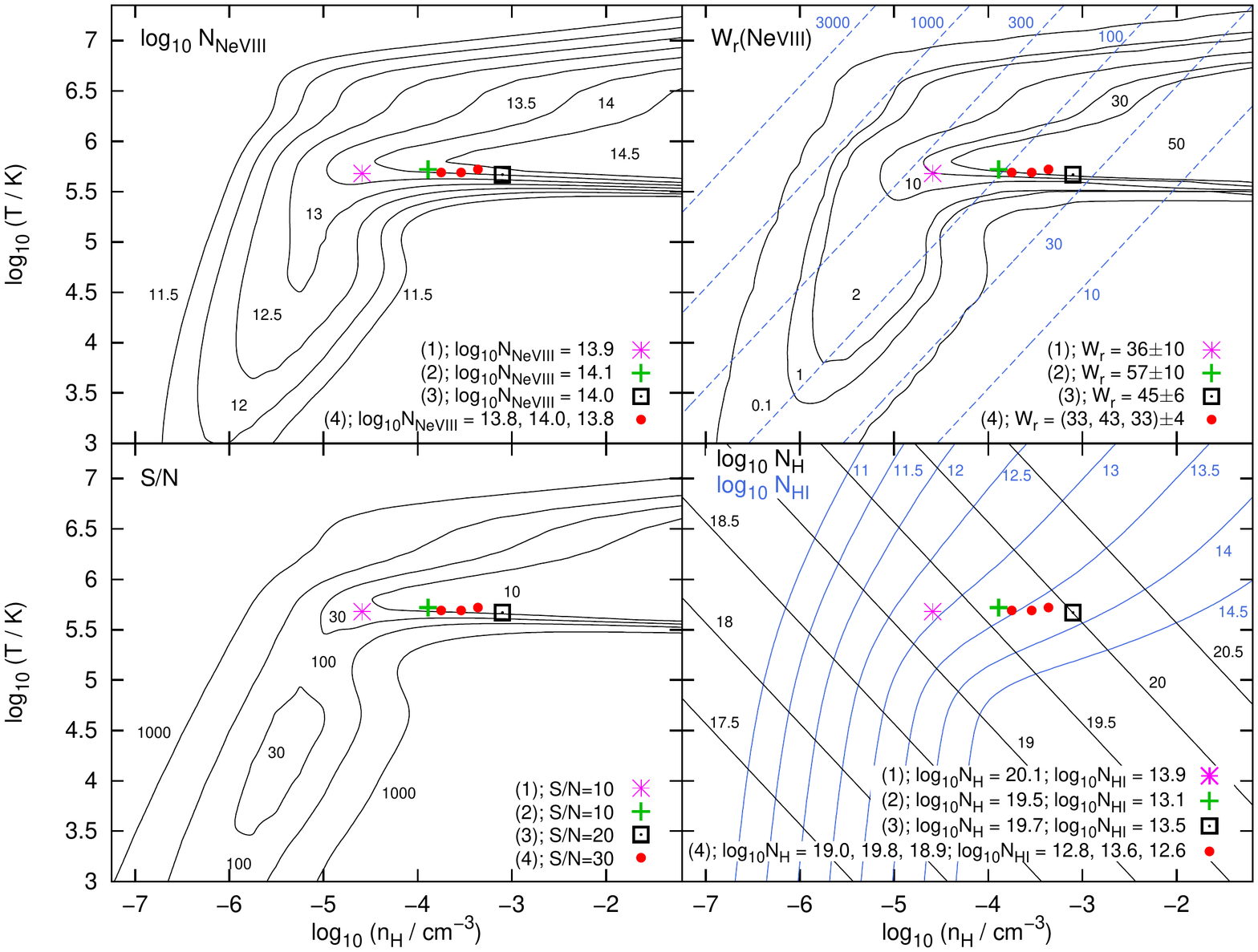}}}
\caption[]{Comparison of the spectral and physical properties predicted by our analytic model to the corresponding measured and inferred properties of a sample of observed \NeVIII\ absorbers at low redshift as reported in the literature: \NeVIII\ column density ($\NNeVIII$; top-left), \NeVIII\ rest-frame equivalent width ($W_r$; top-right; black contours) and linear size in kpc ($l$; top-right; blue dashed lines), total hydrogen column density ($\NH$; bottom-right; black lines) and neutral hydrogen column density ($\NHI$; bottom-right; blue curves). The contours in the bottom-left panel show the minimum sensitivity in terms of S/N required to detect gas at a given density and temperature using resolved \NeVIII\ absorption at 5$\sigma$ significance. The values for each quantity measured (or inferred) from observations are indicated in the legend along with their corresponding reference. A complete list of the quantities measured and inferred from observations is given in Table \ref{tbl:data}. The S/N values quoted in the bottom-left panel correspond to the average S/N of the data in the spectral range of interest. Note that the contours in each panel are identical to the corresponding contours shown in Figure \ref{fig:predict}. Reference key: (1) \citet{sav05a,sav11b}; (2) \citet{nar09b,nar12a}; (3) \citet{nar11a}; (4)  \citet{mei13a}.}
\label{fig:compare}
\end{figure*}
%--------------------------------------------------------------------------------------------------------------------------------------------------------------------------------

Further insight into the nature of \NeVIII\ absorbers at low redshift can be obtained by comparing the distribution of \NeVIII\ mass in gas in our simulation (coloured regions) with the spectral properties of the gas as a function of its density and temperature predicted by our analytic model (contours) shown in each panel. We see that in our model universe most of the \NeVIII\ mass is contained in gas with densities $\log_{10} (\nH / \pcc) \lesssim -4$ (corresponding to overdensities $\Delta \lesssim 150$ at $z=0.5$) and temperatures $T \lesssim 10^{6} \K$ for which the \NeVIII\ column densities are expected to be $\NNeVIII \lesssim 10^{14} \psc$ and the corresponding rest-frame equivalent widths $W_r \lesssim 30$~m\AA. This result is consistent with the distribution of column densities and equivalent widths obtained from our synthetic spectra (Section \ref{sec:stats}). The gas containing most of the \NeVIII\ mass in our simulation thus produces weak to moderate \NeVIII\ absorption features with optical depths at the centre $\log_{10} \tau_0 < -0.5$; therefore, S/N values of 30 or higher are required to detect these features (at 5$\sigma$ significance). The line widths implied by the corresponding gas temperatures, if thermal broadening dominates, are $\bneviii < 30 \kms$, with most of the \NeVIII\ mass  in gas producing thermally broadened lines with $10 \kms < \bneviii < 30 \kms$. The gas responsible for the \NeVIII\ absorption produces \HI\ absorption displaying a wide range of \HI\ column densities: $10^{11} < (\NHI / \psc) < 10^{14}$. The average baryon content of this gas in the form of hydrogen corresponds to total column densities $\log_{10} (\NH / \psc) < 19.5$.

We find that only a small fraction of the \NeVIII\ mass in our simulation resides in gas with densities $\nH \gtrsim 10^{-4} \pcc$ and temperatures around $T \approx 5\times10^5 \K$, i.e.\ in gas which produces \NeVIII\ absorption with column densities $\NNeVIII \gtrsim 10^{14} \psc$ (and $W_r > 30~$m\AA).  These \NeVIII\ absorbers should be detectable in absorption in spectra with S/N~$\gtrsim$~10 (at $5\sigma$ significance), given that their line strength is $\log_{10} \tau_0 \gtrsim -0.5$. The gas temperatures imply line widths $\bneviii \gtrsim 20 \kms$. The average total hydrogen column densities of the gas responsible for this type of \NeVIII\ absorption can be as high as $\NH  \sim 10^{20} \psc$. Due to the high gas temperatures, the hydrogen neutral fraction in this gas is low, and the gas produces \HI\ absorption of BLA type: broad, with $\bhi \gtrsim 90 \kms$, and shallow, with column densities $10^{13} \lesssim (\NHI / \psc) \lesssim 10^{14}$; a high sensitivity is thus required to detect this gas phase using \HI\ absorption. The linear extent of absorbers with densities $\nH \sim 10^{-4} \pcc$ and temperatures $T < 10^6 \K$ is on the order of $10^2$ kpc, and it decreases with density.  Note, however, that the size, and the \H\ and \HI\ column densities of the gas clouds that {\em harbour} the metal cloudlets responsible for the \NeVIII\ absorption are a factor $c_L^{-1}$ higher. It is worth noting that all these properties closely resemble the properties of the \NeVIII\ absorbers at low redshift reported in the literature to date. We will discuss this in more detail in the next section.

The combined results from our analytic model and our simulation imply that \NeVIII\ absorbers with $\NNeVIII \gtrsim 10^{14} \psc$ (and $W_r \gtrsim 30~$m\AA) are not common at low redshift. However, the weakest members in this class of \NeVIII\ absorbers are just strong enough to allow their detection in absorption spectra with S/N $\sim$ 10. The detection of weaker, but more common \NeVIII\ absorbers requires a significantly higher sensitivity S/N~$\gtrsim$~100. Our model thus simultaneously provides a natural explanation for the low detection rate of \NeVIII\ absorbers at low redshift --first noted by \citet{nar12a}--, as well as for the similarity in their strength ($\NNeVIII \sim 10^{14} \psc$) given that the typical sensitivity of common UV spectra is S/N $\sim$ 10 (see Section \ref{sec:compare}).

Equally important, the result from our model that absorbers with $\NNeVIII \gtrsim 10^{14} \psc$ and $W_r > 30~$m\AA\ are only produced in gas with temperatures in a very narrow range around $T \approx 5\times10^5 \K$ strongly supports the idea that {\em this} particular class of \NeVIII\ absorbers is a sensitive probe of the gas temperature \citep{sav05a}. In other words, the detection of \NeVIII\ absorbers with $\NNeVIII \gtrsim 10^{14} \psc$ is a strong indication of the presence of gas with temperatures $\log_{10} (T / \K) \gtrsim 5.5$, {\em under ionisation equilibrium conditions}.

Our model further predicts that the gas that produces such strong \NeVIII\ absorption can have total hydrogen column densities $\NH \gg 10^{19} \psc$. Nevertheless, the overall baryon content of low-redshift \NeVIII\ absorbers may not be as high as previously expected \citep[e.g.][]{sav05a}. Indeed, the total baryon content of the gas producing the \NeVIII\ absorption in our simulation amounts to $\Omega_{b}(\NeVIII) = 1.22\times10^{-3}$ or roughly 3 per cent of the total baryon budget. The gas producing absorbers with $\NNeVIII \gtrsim 10^{14} \psc$ contain less than 0.5 per cent of the total baryons. Note, however, that our estimate represents a strict lower limit, since we ignore absorbers with $\NNeVIII \lesssim 10^{12} \psc$ which contain roughly 20 per cent of the \NeVIII\ mass in our simulation (see Section \ref{sec:stats}), although according to our model predictions their average total hydrogen column densities, and in consequence their baryon content, is expected to be relatively low, $\NH < 10^{18} \psc$ . Note also that the {\em total} baryon density {\em traced} by the \NeVIII\ absorbers is a factor $c_L^{-1}$ higher, or approximately 10 per cent of the total baryon budget in our simulation if $c_L \approx 0.3$. Our result is of the same order as the corresponding value $\Omega_b(\NeVIII) = 2.63\times10^{-3}$ (or 6 per cent of the total baryons in the Universe if $\Omega_b = 0.0418$) estimated by \citet{nar09b}, based on a single detection, and the value $\Omega_b(\NeVIII) \lesssim 2.00\times10^{-3}$ (roughly 4 per cent of the baryon density if $\Omega_b = 0.0418$) obtained by \citet{mei13a} based on three detections. 

On the other hand, if $c_L < 0.3$, then the baryon content traced by \NeVIII\ in our simulation could be significantly higher. For instance, if $c_L \sim 0.1$ as is suggested by observations in combination with our analytic model (see Section \ref{sec:compare}), then the fraction of the total baryon contained in the (warm-hot) gas harbouring the \NeVIII\ would be boosted to roughly 30 per cent, which is comparable to the fraction of baryons in gas traced by BLAs in our simulation \citep[25 per cent; ][]{tep12a}. However, the overlap between BLAs and \NeVIII\ in terms of the baryons traced by each of these ions is unclear, and in consequence the question whether these components together or separately are enough to close the baryon budget of warm-hot diffuse gas at low redshift remains unanswered.

%--------------------------------------------------------------------------------------------------------------------------------------------------------------------------------
\section{Comparison with observations} \label{sec:compare}

Given the relevance of the results presented in the previous section, it is important to test the predictions of our analytic model using the measured (and inferred) properties of \NeVIII\ absorbers observed at low redshift. In Table \ref{tbl:data} we present all the \NeVIII\ detections available in the literature to date, listing only those physical parameters that are relevant for our study (if available). All the values listed in Table \ref{tbl:data} have been taken from the corresponding reference given in the last column. The only exception is the rest-frame equivalent width ($W_r$), which we have computed with the help of equations \eqref{eq:ew_1} and \eqref{eq:tau0_neviii} using the values for the \NeVIII\ column density ($\NNeVIII$) and the Doppler parameter ($\bneviii$) listed on columns 3 and 4.

Note that in some cases \citep[e.g.][]{sav05a,mei13a} the gas temperature of the collisionally ionised gas phase responsible for the \NeVIII\ (and associated \OVI) absorption has been obtained by scaling $\NH$ in order for the model to reproduce the measured \NeVIII\ and \OVI\ column densities, assuming CIE and some value for the metallicity  (usually the same value of the metallicity inferred for the photo-ionised gas phase traced by low ions, e.g., \CIII, \SiIII). In these cases, $\NHI$ has been calculated from the inferred temperature which sets the \HI\ ionisation fraction $\fHI$ (assuming CIE) and the adopted value of $\NH$ using $\NHI = \fHI ~\NH$. As shown by \citet{nar12a}, however, assuming the metallicities of the photo-ionised and collisionally ionised phases to be similar may not be always justified.

In others cases \citep[e.g.][]{nar12a}, the gas temperature of the observed \NeVIII\ absorbing gas has been inferred from the ratio of the observed \NeVIII\ to \OVI\ column densities ($\NNeVIII/\NOVI$) assuming CIE and fixed (usually solar) elemental abundances. The inferred temperature (together with the assumption of CIE) has then been used to convert the neutral hydrogen column density ($\NHI$) into a total hydrogen column density ($\NH$). In these cases, the \HI\ column density of the gas phase responsible for the \NeVIII\ and \OVI\ absorption has been either directly measured in the spectrum \citep[e.g.][]{sav11b} or estimated by overlaying onto the \HI\ absorption feature detected in the observed spectrum an additional \HI\ component whose width is set by the inferred gas temperature and whose \HI\ column density is the largest possible such that the overall fit is still consistent with the data \citep[e.g.][]{nar11a,nar12a}.

With one notable exception, there are no estimates (apart from upper or lower limits) of the density of the \NeVIII\ absorbers available in the literature. The only estimate is given by \citet{nar11a} for a \NeVIII\ absorber at $z \approx 0.5$. They obtain $\log_{10} (\nH / \pcc ) = -3.10$ using a model that combines the effects of photo- and collisional ionisation (as opposed to assuming CIE only) to compute the ionisation balance of the absorbing gas (see Appendix \ref{sec:current} for more details on the analysis of this absorbing system). Given the lack of information on the density of \NeVIII\ in the literature, in the following we assume that the density of the gas responsible for the \NeVIII\ absorption is comparable to the density of the gas giving rise to the absorption by low-ionisation species (e.g. \OIII). It should be mentioned that the density thus estimated is uncertain. First, there is no particular reason why the density of the collisionally ionised gas should be equal to the photo-ionised gas, although it is unlikely that they differ by order(s) of magnitude (see left panel of Figure \ref{fig:phys}). Second, because the density of the photo-ionised gas has been inferred assuming photo-ionisation equilibrium (PIE), which may not be satisfied in general. And last, because its inferred value depends on the intensity of the UV ionising field, which is itself uncertain either because it may locally deviate from the assumed UV background radiation, or because the latter is also uncertain by factors of a few \citep[e.g.][]{haa12a}.

Figure \ref{fig:compare} shows the comparison of the physical parameters predicted by our analytic model to the corresponding measured and inferred parameters of the \NeVIII\ absorbers listed in Table \ref{tbl:data}. Note that we use only the subset of the observations for which we have an estimate of the gas density ($\nH$). The panels in this figure display a series of contours which show, from the top-left and moving clockwise, the \NeVIII\ column density ($\NNeVIII$), the rest-frame \NeVIII\ equivalent width ($W_r$), the total hydrogen column density ($\NH$; black lines) and the neutral hydrogen column density ($\NHI$; blue curves) of the \NeVIII\ absorbing gas, and the minimum required sensitivity (in terms of S/N) to detect gas using \NeVIII\ absorption as a function of the gas density and temperature, assuming the line to be fully resolved. The top-right panel shows additionally the approximate size (in kpc) across the \NeVIII\ absorber (blue dashed lines). Note that the contours are identical to the corresponding contours shown in Figure \ref{fig:predict}. The data points represent the observed \NeVIII\ absorbers. In each panel, the measured (or inferred) value of a given quantity (e.g.\ $\NNeVIII$) is indicated in the legend, along with the reference it has been taken from (see the figure caption and Table \ref{tbl:data}).

We find broad agreement between  the results inferred from observations and the predictions of our analytic model. The panels in the top row, for instance, demonstrate that the measured \NeVIII\ column densities, as well as the measured rest-frame \NeVIII\ equivalent widths, of the observed \NeVIII\ absorbers with a given inferred density and inferred temperature fall within the range of \NeVIII\ column densities and equivalent widths predicted by our model for gas in the corresponding density and temperature range. It is remarkable that all the observed \NeVIII\ absorbers have very similar strengths, $\NNeVIII \sim 10^{14} \psc$, within the given uncertainties, and that they have all been detected in spectra with  (average) S/N~$\sim$~10, which is perfectly consistent with the result from our model (compare top-left and bottom-left panels). The bottom-right panel shows that both the ranges of \H\ and \HI\ column densities of the gas responsible for the \NeVIII\ absorption obtained from observations are  well predicted by our analytic model. With the exception of the \NeVIII\ absorber reported by \citet[][magenta star]{sav05a}, our model predicts that the typical linear extension of the absorbers is on the order $10 - 100 ~\kpc$ (see top-right panel). Note that these values are consistent with the values that result if we estimate the linear size from the relation $l = \NH / \nH$, using the values listed in Table \ref{tbl:data}.

It is noteworthy that all the measured properties of the $z=0.495$ \NeVIII\ absorber detected by \citet[][black square]{nar11a} are very well reproduced by our analytic model, even though the neon abundance in this absorbers is inferred to be higher ($[\Ne / \H ] = -0.44$) than our adopted value ($[\Ne / \H ] = -0.7$). In contrast, our model fails in reproducing the observed properties (in particular $l$, $\NH$ and $\NHI$) of the \NeVIII\ absorber detected by \citet[][magenta star]{sav05a}. We note with interest, however, that the disagreement with our model predictions can be resolved if the density of this absorber is assumed to be significantly higher. Indeed, if this absorber had a density $\nH \sim 10^{-3} \pcc$ rather than $\nH \sim 10^{-5} \pcc$, then the values predicted by our model would closely match each of its corresponding observed properties, yielding in particular a more realistic size $l \approx 30$~kpc (rather than $300 ~\kpc$) if given by $l = \NH / \nH$. This suggests that our assumption that the density of the photo- and collisionally are comparable might not justified in this case.

Of particular interest are the \NeVIII\ absorbers reported by \citet[][red dots]{mei13a}. Assuming that the close alignment in velocity space of the \NeVIII\ absorbing collisionally ionised gas with the cooler, photo-ionised gas traced by e.g., \OIII, implies their coincidence in physical space, \citet{mei13a} rule out that the \NeVIII\ absorbing gas be either pressure confined or in hydrostatic equilibrium with the photo-ionised gas. They suggest that the \NeVIII\ absorption arises in a conductive interface between the cool gas and the hot halo gas. Interestingly, although the absorbing properties of such an interface are not captured by our model, our predictions are broadly consistent with the observed properties of these \NeVIII\ absorbers. It has been argued \citep[e.g.][]{sav11b} that an unrealistically large number of such interfaces is needed to produce the observed column densities of \OVI. To date, no such analysis has been conducted for \NeVIII. However, since \OVI\ is always detected whenever \NeVIII\ is, it is reasonable to assume that the same argument applies to \NeVIII. Hence, if and how many of the observed \NeVIII\ absorbers can be explained by such a scenario remains an open question.

The fact that the average neon abundance of the observed \NeVIII\ absorbers ($[\Ne / \H] \approx -0.2$; see Figure \ref{fig:meta}) is higher by 0.5 dex than the fiducial value of our analytic model has an interesting implication. As discussed in Section \ref{sec:param}, consistency between our analytic model and our simulation demands that $c_L$ decreases if $[\Ne /\H]$ increases. More precisely, increasing the value of $[\Ne /\H]$ by 0.5 dex higher would require us to decrease $\log_{10} c_L$ by the same amount. Therefore, adopting $[\Ne / \H] = -0.2$ would require $\log_{10} c_L = -1$. In other words, the typical neon abundance observed in low redshift \NeVIII\ absorbers, in combination with our analytic model, favours values $c_L \sim 10^{-1}$, thus supporting the idea that metals are in-homogeneously distributed at low redshift as well. Such a change in the values of $[\Ne /\H]$ and $c_L$ would leave the predicted values for $\NNeVIII$ and $W_r$ (top-row panels of Figure \ref{fig:compare}) unaltered, but it would reduce the predicted sizes $l$, and the gas column densities $\NH$ and $\NHI$ by a factor $10^{\,0.5} \approx 3$, all of which would still be consistent with the results inferred from observations. Note that, in particular, the predicted sizes of absorbers with $\NNeVIII \sim 10^{14} \psc$ would be on the of order 10 kpc.

It is noteworthy that the inferred gas temperature of all observed \NeVIII\ absorbers is very similar and close to the temperature $T \approx 5\times10^5 \K$ at which the \NeVIII\ fraction peaks in CIE. In contrast, their densities span almost two orders of magnitude. However, if we ignore the  \NeVIII\ absorber detected by \cite{sav05a} (or assume this absorbers has a density $\nH \sim 10^{-3} \pcc$ as discussed above), then all \NeVIII\ absorbers observed at low redshift appear to be produced under rather uniform physical conditions,  that is in gas with $\nH \sim 10^{-4} - 10^{-3} \pcc$ and $T \approx 5\times10^5 \K$, and total hydrogen column densities $\NH \sim 10^{19} - 10^{20} \psc$. The relatively high densities of the observed \NeVIII\ absorbers, which correspond to overdensities of $\Delta \approx 1.5\times(10^2 - 10^3)$ at $z=0.5$, indicate that they do not originate in the diffuse, warm-hot intergalactic medium but rather in the vicinity of galaxies. This is consistent with the detection by \citet{mul09a} of an  $L \approx 0.25 L^*$ galaxy at an impact parameter $\rho \leq 200 \hkpc$ (physical) from the \los\ toward HE~0226-4110 within $| \Delta v | \leq 300 \kms$ of the $z=0.207$ \NeVIII\ absorber detected by \citet{sav05a}.\\

It is important to stress that we do not expect a perfect agreement between our analytic model and the observations for several reasons. First, an accurate comparison of the predicted and measured column densities and equivalent widths is difficult, given that our model assumes the lines to be fully resolved while the intrinsic velocity structure of the observed lines may be blurred by the limited resolution and sensitivity (S/N) of the data, as  demonstrated for instance in Figure \ref{fig:spec}. Second, the physical parameters of the \NeVIII\ absorbers obtained from observations are based on models that assume collisional ionisation equilibrium \citep[but see][]{nar11a}, while we take both the effect both of photo- and collisional ionisation into account to compute the ionisation balance of the absorbing gas (although we also assume ionisation equilibrium). Furthermore, the gas density ($\nH$) of the observed \NeVIII\ absorbers, which we have assumed to be comparable to the density of the associated photo-ionised gas phase may not be justified in general. Finally, our analytic model assumes local hydrostatic equilibrium \citep{sch01a} which may hold only approximately and not necessarily for all absorbers.

In spite of these uncertainties, we find that our analytic model broadly reproduces the properties of observed \NeVIII\ absorbers at low redshift, although the number of current detections is still small. Thus, more data is required before we can test the predictions of our model in a more robust way. Such data should soon become available from the analysis of a large sample of high-quality spectra obtained with COS.\\

The results presented above lead us to conclude that the intervening \NeVIII\ absorbers reported to date in the literature likely correspond to collisionally ionised, baryon-rich (in terms of $\NH$, with $\NH \sim 10^{19} - 10^{20} \psc$), compact ($\sim 10 ~\kpc$) metal absorbers produced in gas with temperatures well above $10^5 \K$ and with densities $\nH \gg 10^{-5} \pcc$. These strong ($\NNeVIII \sim 10^{14} \psc$) \NeVIII\ absorbers most probably originate close to galaxies, and they make up only a  small fraction of the general \NeVIII\ absorber population at low redshift. The latter consists mainly of weaker \NeVIII\ absorbers with $\NNeVIII < 10^{14} \psc$, which are produced in diffuse gas with $\nH \sim 10^{-5}$ and hydrogen column densities $\NH \lesssim 10^{19} \psc$, and whose detection requires a sensitivity in terms of S/N on the order 100.

%--------------------------------------------------------------------------------------------------------------------------------------------------------------------------------
\section{Summary and conclusions} \label{sec:sum}

In recent years a substantial fraction of the ultra-violet (UV) spectroscopy of quasars (QSO) has been dedicated to tracking down a peculiar gas phase characterised by its low densities ($\nH \sim 10^{-6} - 10^{-5} \pcc$) and relatively high temperatures ($T > 10^5 \K$) which is expected to harbour a substantial fraction of the baryons in the Universe at low redshift. Next to other ions such as five-times ionised oxygen (\OVI) and broad \HI\ \lya\ absorbers (BLAs), seven-times ionised neon (\NeVIII) has been considered to be a potential tracer of this gas phase, especially because of the temperature ($T \approx 5\times10^5 \K$) at which the \NeVIII\ ionisation fraction is highest in collisional ionisation equilibrium. But despite the relatively high abundance of neon in the Universe, there have so far only been a small number of \NeVIII\ detections at low redshift, and consequently the physical conditions and the fraction of the baryons traced by the gas phase producing the observed \NeVIII\ absorption are still uncertain.

In this paper we have investigated the physical state of the \NeVIII\ absorbing gas at low redshift following two independent but complementary  approaches. First, we have used a large cosmological simulation of structure formation which includes many of the physical processes relevant to the production and distribution of metals in the Universe. Second, we have developed an analytic model to predict the  signatures of \NeVIII\ absorbing gas as a function of the gas density, temperature, and the abundance of metals in the gas. The assumptions underlying our model are that the metals are in-homogeneously distributed within the gaseous structures responsible for the absorption, that these structures are in quasi hydrostatic equilibrium with sizes of the order of the local Jeans length, that the gas is only exposed to the UV/X-ray background, and that it is in ionisation equilibrium. The last two assumptions were also used in the simulation. We quantify the degree of metal mixing by introducing a parameter, $c_L$, which corresponds to the fraction of the length of the gas cloud along the line of sight that intersects cloudlets enriched with neon, where the enriched cloudlets have a similar density and temperature as the surrounding gas. We have shown that the simulation and the analytic model are consistent with each other, and that the predictions from our analytic model are broadly consistent with the results from observations, provided that $c_L \sim 10^{-1}$, indicating that metals are poorly mixed with the ambient gas.

The most important results we obtain from the combination of our simulation and our analytic model regarding the physical state of the \NeVIII\ absorbing gas at redshift $z=0.5$ are the following:

\begin{enumerate}

\item High \NeVIII\ ion fractions ($\fNeVIII \sim 0.1$) can be produced by both photo- and collisional ionisation, but most of the \NeVIII\ in our simulation is found in collisionally ionised gas with densities $\nH \sim 10^{-5} \pcc$ and $T \approx  5\times10^5 \K$ (Section \ref{sec:sim}).

\item\ \NeVIII\ absorption with column densities $\NNeVIII \gtrsim 10^{14} \psc$ ($W_r \gtrsim 30~$m\AA) is only produced in collisionally ionised gas with temperatures around $T = 5\times10^5 \K$ and $\nH \gtrsim 10^{-4} \pcc$, total hydrogen column densities $\NH \sim 10^{20} \psc$, and neutral hydrogen column densities $\NHI  \sim 10^{13} - 10^{14} \psc$. The detection of such absorbers is therefore a strong indication of the presence of gas with $T > 10^5 \K$, provided the gas is in ionisation equilibrium (Section \ref{sec:results}).

\item At $z=0.5$ the number of strong ($W_r \gtrsim 30~$m\AA\ or $\NNeVIII \gtrsim 10^{14} \psc$) \NeVIII\ absorbers per unit redshift is predicted to be $dN/dz \sim 10^{-1}$, which is low compared to the corresponding value inferred from observations: $dN/dz \sim 1$. If real, this discrepancy may indicate that the strong \NeVIII\ absorption has been boosted by non-equilibrium effects due to past AGN activity \citep{opp13b}. In our simulation, the dominant population of  \NeVIII\ absorbers at low redshift consists of weaker absorbers with $\NNeVIII < 10^{14} \psc$ whose detection requires S/N $\gtrsim$ 100 (Sections \ref{sec:stats}, \ref{sec:results}). 

\item At a given \NeVIII\ column density, the linear extent of the absorbers scales inversely proportional to the metal abundance. Metal-rich ($[\Ne / \H] \approx 0$), strong ($\NNeVIII \gtrsim 10^{14} \psc$) absorbers have typical sizes $\sim 10~\kpc$, provided that $c_L \sim 10^{-1}$ (Sections \ref{sec:phys}, \ref{sec:param}, \ref{sec:compare}). The gas clouds harbouring these \NeVIII\ cloudlets are, however, a factor $c_L^{-1}$ larger.

\item The gas producing the \NeVIII\ absorption in our simulations contains roughly 3 per cent of the total amount of baryons; absorbers with $\NNeVIII \gtrsim 10^{14} \psc$ contain less than one per cent of the total baryons in our simulation. The baryon content of the gas harbouring the \NeVIII\ absorbers (and which has a similar density and temperature as the \NeVIII\ cloudlets) is in each case a factor $c_L^{-1}$ higher (Section \ref{sec:results}).

\item The neon abundances of the \NeVIII\ absorbing gas in our simulation are distributed in the range $-1.5 \lesssim [\Ne/\H] \lesssim 0$, with  median $[\Ne / \H ] = -0.7$ (Section \ref{sec:param}).

\item The distribution of \NeVIII\ column densities obtained from synthetic spectra at $z=0.5$ is well described by a single power law with index $\beta = 2.9 \pm 0.1$, in the range $\log_{10} (\NNeVIII / \psc) \in [12.5, 14.5]$ (Section \ref{sec:stats}).

\item The small number of \NeVIII\ absorbers observed at low redshift to date appear to belong to a class of scarce, strong ($\NNeVIII \sim 10^{14} \psc$), baryon-rich ($\NH \sim 10^{20} \psc$), compact ($\sim10 ~\kpc$) metal absorbers produced in gas with densities $\nH \sim 10^{-4} - 10^{-3} \pcc$ and temperatures $T \approx 5\times10^5 \K$. The relatively high densities of these \NeVIII\ absorbers indicate that they likely originate in the immediate vicinity of galaxies, rather than in the diffuse, warm-hot intergalactic medium (Section \ref{sec:compare}).\\

We conclude that strong \NeVIII\ absorbers are robust probes of shock-heated diffuse gas. These absorbers do not, however, contain significant amounts of baryons as a consequence of the poor mixing of metals. The current low detection rate of \NeVIII\ absorbers at low redshift is a result of the limited sensitivity of current UV spectra (S/N $\sim$ 10). Spectra with S/N~$\sim 100$ would allow the detection of a larger number of weaker systems embedded in structures that harbour a substantial fraction ($\gtrsim 10$ per cent) of the baryons at low redshift.  

\end{enumerate}

%--------------------------------------------------------------------------------------------------------------------------------------------------------------------------------
\section*{Acknowledgements}

We thank the anonymous reviewer for a very constructive report. We are also grateful to all members of the OWLS team. We acknowledge in particular the contributions of Craig M.~Booth and Tom Theuns to \textsc{specwizard} and the help the former provided in the use of the simulations. The simulations presented here were run on Stella, the LOFAR Blue Gene/L system in Groningen and on the Cosmology Machine at the Institute for Computational Cosmology in Durham as part of the Virgo Consortium research programme. This work was sponsored by the National Computing Facilities Foundation (NCF) for the use of supercomputer facilities, with financial support from the Netherlands Organisation for Scientific Research (NWO), an NWO VIDI grant, the Marie Curie Initial Training Network CosmoComp (PITN-GA-2009-238356), the {\em Deutsche Forschungsgemeinschaft} (DFG) through Grant DFG-GZ: Ri 1124/5-1, and the European Research Council under the European Unions Seventh Framework Programme (FP7/2007-2013) / ERC Grant agreement 278594-GasAroundGalaxies.\\

%--------------------------------------------------------------------------------------------------------------------------------------------------------------------------------
% For the submission (both to journal and arXiv) comment out the 3 following lines:
%\bibliographystyle{/Users/tepper/latex/mnras/mn2e_eprint} % style mn2e_eprint.bst, allows eprint field to be included
%\bibliographystyle{/Users/tepper/latex/mnras/mn2e} % style mn2e.bst
%\bibliography{/Users/tepper/references/complete} % references list

\begin{thebibliography}{}

\bibitem[\protect\citeauthoryear{{Altay}, {Theuns}, {Schaye}, {Crighton} \&
  {Dalla Vecchia}}{{Altay} et~al.}{2011}]{alt11a}
{Altay} G.,  {Theuns} T.,  {Schaye} J.,  {Crighton} N.~H.~M.,    {Dalla
  Vecchia} C.,  2011, \apjl, 737, L37+, \eprint{1012.4014}

\bibitem[\protect\citeauthoryear{{Anders} \& {Grevesse}}{{Anders} \&
  {Grevesse}}{1989}]{and89a}
{Anders} E.,  {Grevesse} N.,  1989, \gca, 53, 197

\bibitem[\protect\citeauthoryear{Bertone, Schaye \& Dolag}{Bertone
  et~al.}{2008}]{ber08a}
Bertone S.,  Schaye J.,    Dolag K.,  2008, Space Science Reviews, 134, 295

\bibitem[\protect\citeauthoryear{{Booth} \& {Schaye}}{{Booth} \&
  {Schaye}}{2009}]{boo09a}
{Booth} C.~M.,  {Schaye} J.,  2009, \mnras, 398, 53, \eprint{0904.2572}

\bibitem[\protect\citeauthoryear{Booth \& Schaye}{Booth \&
  Schaye}{2011}]{boo11a}
Booth C.~M.,  Schaye J.,  2011, \mnras, 413, 1158

\bibitem[\protect\citeauthoryear{{Cen}}{{Cen}}{2012}]{cen12a}
{Cen} R.,  2012, \apj, 753, 17, \eprint{1112.4527}

\bibitem[\protect\citeauthoryear{{Cen} \& {Ostriker}}{{Cen} \&
  {Ostriker}}{1999}]{cen99a}
{Cen} R.,  {Ostriker} J.~P.,  1999, \apj, 514, 1, \eprint{9806281}

\bibitem[\protect\citeauthoryear{{Dalla Vecchia} \& {Schaye}}{{Dalla Vecchia}
  \& {Schaye}}{2008}]{dal08b}
{Dalla Vecchia} C.,  {Schaye} J.,  2008, \mnras, 387, 1431, \eprint{0801.2770}

\bibitem[\protect\citeauthoryear{{Danforth} \& {Shull}}{{Danforth} \&
  {Shull}}{2008}]{dan08b}
{Danforth} C.~W.,  {Shull} J.~M.,  2008, \apj, 679, 194, \eprint{0709.4030}

\bibitem[\protect\citeauthoryear{{Danforth}, {Shull}, {Rosenberg} \&
  {Stocke}}{{Danforth} et~al.}{2006}]{dan06a}
{Danforth} C.~W.,  {Shull} J.~M.,  {Rosenberg} J.~L.,    {Stocke} J.~T.,  2006,
  \apj, 640, 716, \eprint{astro-ph/0508656}

\bibitem[\protect\citeauthoryear{{Danforth}, {Stocke} \& {Shull}}{{Danforth}
  et~al.}{2010}]{dan10a}
{Danforth} C.~W.,  {Stocke} J.~T.,    {Shull} J.~M.,  2010, \apj, 710, 613,
  \eprint{0912.1603}

\bibitem[\protect\citeauthoryear{{Dav{\'e}}, {Cen}, {Ostriker}, {Bryan},
  {Hernquist}, {Katz}, {Weinberg}, {Norman} \& {O'Shea}}{{Dav{\'e}}
  et~al.}{2001}]{dav01a}
{Dav{\'e}} R.,  {Cen} R.,  {Ostriker} J.~P.,  {Bryan} G.~L.,  {Hernquist} L.,
  {Katz} N.,  {Weinberg} D.~H.,  {Norman} M.~L.,    {O'Shea} B.,  2001, \apj,
  552, 473, \eprint{astro-ph/0007217}

\bibitem[\protect\citeauthoryear{{Dav{\'e}}, {Hernquist}, {Weinberg} \&
  {Katz}}{{Dav{\'e}} et~al.}{1997}]{dav97a}
{Dav{\'e}} R.,  {Hernquist} L.,  {Weinberg} D.~H.,    {Katz} N.,  1997, \apj,
  477, 21, \eprint{astro-ph/9609115}

\bibitem[\protect\citeauthoryear{{Ferland}, {Korista}, {Verner}, {Ferguson},
  {Kingdon} \& {Verner}}{{Ferland} et~al.}{1998}]{fer98a}
{Ferland} G.~J.,  {Korista} K.~T.,  {Verner} D.~A.,  {Ferguson} J.~W.,
  {Kingdon} J.~B.,    {Verner} E.~M.,  1998, \pasp, 110, 761

\bibitem[\protect\citeauthoryear{{Ford}, {Oppenheimer}, {Dav{\'e}}, {Katz},
  {Kollmeier} \& {Weinberg}}{{Ford} et~al.}{2013}]{for13a}
{Ford} A.~B.,  {Oppenheimer} B.~D.,  {Dav{\'e}} R.,  {Katz} N.,  {Kollmeier}
  J.~A.,    {Weinberg} D.~H.,  2013, \mnras

\bibitem[\protect\citeauthoryear{{Fukugita}}{{Fukugita}}{2004}]{fuk04a}
{Fukugita} M.,  2004, in {Ryder} S.,  {Pisano} D.,  {Walker} M.,   {Freeman}
  K.,  eds, Dark Matter in Galaxies Vol.~220 of IAU Symposium, {Cosmic Matter
  Distribution: Cosmic Baryon Budget Revisited}.
pp 227--+

\bibitem[\protect\citeauthoryear{{Green}, {Froning}, {Osterman}, {Ebbets},
  {Heap}, {Leitherer}, {Linsky} \& {Savage} B.~D.}{{Green}
  et~al.}{2012}]{gre12a}
{Green} J.~C.,  {Froning} C.~S.,  {Osterman} S.,  {Ebbets} D.,  {Heap} S.~H.,
  {Leitherer} C.,  {Linsky} J.~L.,    {Savage} B.~D. e.~a.,  2012, \apj, 744,
  60, \eprint{1110.0462}

\bibitem[\protect\citeauthoryear{{Green} \& {Morse}}{{Green} \&
  {Morse}}{1998}]{gre98a}
{Green} J.~C.,  {Morse} J.~A.,  1998, Space Telesc.~Sci.~Inst., Newsl.,
  Vol.~15, No.~1, p.~6 - 7, 21, 15, 6

\bibitem[\protect\citeauthoryear{{Haardt} \& {Madau}}{{Haardt} \&
  {Madau}}{2001}]{haa01a}
{Haardt} F.,  {Madau} P.,  2001, in {D.~M.~Neumann \& J.~T.~V.~Tran} ed.,
  Clusters of Galaxies and the High Redshift Universe Observed in X-rays
  {Modelling the UV/X-ray cosmic background with CUBA}.
CEA, Saclay, p. p. 64

\bibitem[\protect\citeauthoryear{{Haardt} \& {Madau}}{{Haardt} \&
  {Madau}}{2012}]{haa12a}
{Haardt} F.,  {Madau} P.,  2012, \apj, 746, 125, \eprint{1105.2039}

\bibitem[\protect\citeauthoryear{{Jarosik}, {Bennett}, {Dunkley}, {Gold},
  {Greason}, {Halpern}, {Hill} \& {Hinshaw}}{{Jarosik} et~al.}{2011}]{jar11a}
{Jarosik} N.,  {Bennett} C.~L.,  {Dunkley} J.,  {Gold} B.,  {Greason} M.~R.,
  {Halpern} M.,  {Hill} R.~S.,    {Hinshaw} e.~a.,  2011, \apjs, 192, 14,
  \eprint{1001.4744}

\bibitem[\protect\citeauthoryear{Kramida \& Buchet-Poulizac}{Kramida \&
  Buchet-Poulizac}{2006}]{kra06a}
Kramida A.~E.,  Buchet-Poulizac M.-C.,  2006, The European Physical Journal D -
  Atomic, Molecular, Optical and Plasma Physics, 38, 265

\bibitem[\protect\citeauthoryear{{Lehner}, {Savage}, {Richter}, {Sembach},
  {Tripp} \& {Wakker}}{{Lehner} et~al.}{2007}]{leh07a}
{Lehner} N.,  {Savage} B.~D.,  {Richter} P.,  {Sembach} K.~R.,  {Tripp} T.~M.,
    {Wakker} B.~P.,  2007, \apj, 658, 680, \eprint{0612275}

\bibitem[\protect\citeauthoryear{Levenberg}{Levenberg}{1944}]{lev44a}
Levenberg K.,  1944, Q. Appl. Math., 2, 164

\bibitem[\protect\citeauthoryear{Marquardt}{Marquardt}{1963}]{mar63a}
Marquardt D.,  1963, J. Soc. Ind. Appl. Math., 11, 431

\bibitem[\protect\citeauthoryear{{McCarthy}, {Schaye}, {Ponman}, {Bower},
  {Booth}, {Dalla Vecchia}, {Crain}, {Springel}, {Theuns} \&
  {Wiersma}}{{McCarthy} et~al.}{2010}]{mcc10a}
{McCarthy} I.~G.,  {Schaye} J.,  {Ponman} T.~J.,  {Bower} R.~G.,  {Booth}
  C.~M.,  {Dalla Vecchia} C.,  {Crain} R.~A.,  {Springel} V.,  {Theuns} T.,
  {Wiersma} R.~P.~C.,  2010, \mnras, 406, 822, \eprint{0911.2641}

\bibitem[\protect\citeauthoryear{{Meiring}, {Tripp}, {Werk}, {Howk}, {Jenkins},
  {Prochaska}, {Lehner} \& {Sembach}}{{Meiring} et~al.}{2013}]{mei13a}
{Meiring} J.~D.,  {Tripp} T.~M.,  {Werk} J.~K.,  {Howk} J.~C.,  {Jenkins}
  E.~B.,  {Prochaska} J.~X.,  {Lehner} N.,    {Sembach} K.~R.,  2013, \apj,
  767, 49, \eprint{1201.0939}

\bibitem[\protect\citeauthoryear{Mohr, Taylor \& Newell}{Mohr
  et~al.}{2012}]{moh12a}
Mohr P.~J.,  Taylor B.~N.,    Newell D.~B.,  2012, Rev. Mod. Phys., 84, 1527

\bibitem[\protect\citeauthoryear{{Mulchaey} \& {Chen}}{{Mulchaey} \&
  {Chen}}{2009}]{mul09a}
{Mulchaey} J.~S.,  {Chen} H.-W.,  2009, \apjl, 698, L46, \eprint{0905.1327}

\bibitem[\protect\citeauthoryear{{Muzahid}, {Srianand}, {Savage}, {Narayanan},
  {Mohan} \& {Dewangan}}{{Muzahid} et~al.}{2012}]{muz12a}
{Muzahid} S.,  {Srianand} R.,  {Savage} B.~D.,  {Narayanan} A.,  {Mohan} V.,
  {Dewangan} G.~C.,  2012, \mnras, 424, L59, \eprint{1203.3049}

\bibitem[\protect\citeauthoryear{{Narayanan}, {Savage} \& {Wakker}}{{Narayanan}
  et~al.}{2012}]{nar12a}
{Narayanan} A.,  {Savage} B.~D.,    {Wakker} B.~P.,  2012, \apj, 752, 65,
  \eprint{1204.3951}

\bibitem[\protect\citeauthoryear{{Narayanan}, {Savage}, {Wakker}, {Danforth},
  {Yao}, {Keeney}, {Shull}, {Sembach}, {Froning} \& {Green}}{{Narayanan}
  et~al.}{2011}]{nar11a}
{Narayanan} A.,  {Savage} B.~D.,  {Wakker} B.~P.,  {Danforth} C.~W.,  {Yao} Y.,
   {Keeney} B.~A.,  {Shull} J.~M.,  {Sembach} K.~R.,  {Froning} C.~S.,
  {Green} J.~C.,  2011, \apj, 730, 15, \eprint{1008.3192}

\bibitem[\protect\citeauthoryear{{Narayanan}, {Wakker} \& {Savage}}{{Narayanan}
  et~al.}{2009}]{nar09b}
{Narayanan} A.,  {Wakker} B.~P.,    {Savage} B.~D.,  2009, \apj, 703, 74,
  \eprint{0907.3479}

\bibitem[\protect\citeauthoryear{{Oppenheimer} \& {Dav{\'e}}}{{Oppenheimer} \&
  {Dav{\'e}}}{2009}]{opp09b}
{Oppenheimer} B.~D.,  {Dav{\'e}} R.,  2009, \mnras, 395, 1875,
  \eprint{0806.2866}

\bibitem[\protect\citeauthoryear{{Oppenheimer}, {Dav{\'e}}, {Katz}, {Kollmeier}
  \& {Weinberg}}{{Oppenheimer} et~al.}{2012}]{opp12a}
{Oppenheimer} B.~D.,  {Dav{\'e}} R.,  {Katz} N.,  {Kollmeier} J.~A.,
  {Weinberg} D.~H.,  2012, \mnras, 420, 829, \eprint{1106.1444}

\bibitem[\protect\citeauthoryear{{Oppenheimer} \& {Schaye}}{{Oppenheimer} \&
  {Schaye}}{2013a}]{opp13b}
{Oppenheimer} B.~D.,  {Schaye} J.,  2013a, \mnras, 434, 1063,
  \eprint{1303.0019}

\bibitem[\protect\citeauthoryear{{Oppenheimer} \& {Schaye}}{{Oppenheimer} \&
  {Schaye}}{2013b}]{opp13a}
{Oppenheimer} B.~D.,  {Schaye} J.,  2013b, \mnras, 434, 1043,
  \eprint{1302.5710}

\bibitem[\protect\citeauthoryear{{Persic} \& {Salucci}}{{Persic} \&
  {Salucci}}{1992}]{per92a}
{Persic} M.,  {Salucci} P.,  1992, \mnras, 258, 14P,
  \eprint{arXiv:astro-ph/0502178}

\bibitem[\protect\citeauthoryear{{Press}, {Teukolsky}, {Vetterling} \&
  {Flannery}}{{Press} et~al.}{1992}]{num92a}
{Press} W.~H.,  {Teukolsky} S.~A.,  {Vetterling} W.~T.,    {Flannery} B.~P.,
  1992, Numerical Recipes in Fortran 77, second edn.
Cambridge University Press

\bibitem[\protect\citeauthoryear{{Prochaska}, {Weiner}, {Chen}, {Mulchaey} \&
  {Cooksey}}{{Prochaska} et~al.}{2011}]{pro11a}
{Prochaska} J.~X.,  {Weiner} B.,  {Chen} H.-W.,  {Mulchaey} J.,    {Cooksey}
  K.,  2011, \apj, 740, 91, \eprint{1103.1891}

\bibitem[\protect\citeauthoryear{{Rahmati}, {Pawlik}, {Rai{\v c}evi{\'c}} \&
  {Schaye}}{{Rahmati} et~al.}{2013}]{rah13c}
{Rahmati} A.,  {Pawlik} A.~H.,  {Rai{\v c}evi{\'c}} M.,    {Schaye} J.,  2013,
  \mnras, 430, 2427, \eprint{1210.7808}

\bibitem[\protect\citeauthoryear{{Rauch}, {Haehnelt} \& {Steinmetz}}{{Rauch}
  et~al.}{1997}]{rau97a}
{Rauch} M.,  {Haehnelt} M.~G.,    {Steinmetz} M.,  1997, \apj, 481, 601,
  \eprint{astro-ph/9609083}

\bibitem[\protect\citeauthoryear{{Richter}, {Savage}, {Sembach} \&
  {Tripp}}{{Richter} et~al.}{2006}]{ric06a}
{Richter} P.,  {Savage} B.~D.,  {Sembach} K.~R.,    {Tripp} T.~M.,  2006, \aap,
  445, 827, \eprint{astro-ph/0509539}

\bibitem[\protect\citeauthoryear{{Richter}, {Savage}, {Tripp} \&
  {Sembach}}{{Richter} et~al.}{2004}]{ric04a}
{Richter} P.,  {Savage} B.~D.,  {Tripp} T.~M.,    {Sembach} K.~R.,  2004,
  \apjs, 153, 165, \eprint{0403513}

\bibitem[\protect\citeauthoryear{{Savage}, {Lehner} \& {Narayanan}}{{Savage}
  et~al.}{2011}]{sav11b}
{Savage} B.~D.,  {Lehner} N.,    {Narayanan} A.,  2011, \apj, 743, 180,
  \eprint{1111.1697}

\bibitem[\protect\citeauthoryear{Savage, Lehner, Wakker, Sembach,  \&
  Tripp}{Savage et~al.}{2005}]{sav05a}
Savage B.~D.,  Lehner N.,  Wakker B.~P.,  Sembach K.~R.,     Tripp T.~M.,
  2005, The Astrophysical Journal, 626, 776

\bibitem[\protect\citeauthoryear{{Savage} \& {Sembach}}{{Savage} \&
  {Sembach}}{1991}]{sav91a}
{Savage} B.~D.,  {Sembach} K.~R.,  1991, \apj, 379, 245

\bibitem[\protect\citeauthoryear{{Schaye}}{{Schaye}}{2001}]{sch01a}
{Schaye} J.,  2001, \apj, 559, 507, \eprint{astro-ph/0104272}

\bibitem[\protect\citeauthoryear{{Schaye}, {Aguirre}, {Kim}, {Theuns}, {Rauch}
  \& {Sargent}}{{Schaye} et~al.}{2003}]{sch03a}
{Schaye} J.,  {Aguirre} A.,  {Kim} T.,  {Theuns} T.,  {Rauch} M.,    {Sargent}
  W.~L.~W.,  2003, \apj, 596, 768, \eprint{0306469}

\bibitem[\protect\citeauthoryear{{Schaye}, {Carswell} \& {Kim}}{{Schaye}
  et~al.}{2007}]{sch07a}
{Schaye} J.,  {Carswell} R.~F.,    {Kim} T.,  2007, \mnras, 379, 1169,
  \eprint{astro-ph/0701761}

\bibitem[\protect\citeauthoryear{{Schaye} \& {Dalla Vecchia}}{{Schaye} \&
  {Dalla Vecchia}}{2008}]{sch08e}
{Schaye} J.,  {Dalla Vecchia} C.,  2008, \mnras, 383, 1210, \eprint{0709.0292}

\bibitem[\protect\citeauthoryear{{Schaye}, {Dalla Vecchia}, {Booth}, {Wiersma},
  {Theuns}, {Haas}, {Bertone}, {Duffy}, {McCarthy} \& {van de Voort}}{{Schaye}
  et~al.}{2010}]{sch10a}
{Schaye} J.,  {Dalla Vecchia} C.,  {Booth} C.~M.,  {Wiersma} R.~P.~C.,
  {Theuns} T.,  {Haas} M.~R.,  {Bertone} S.,  {Duffy} A.~R.,  {McCarthy} I.~G.,
     {van de Voort} F.,  2010, \mnras, 402, 1536, \eprint{0909.5196}

\bibitem[\protect\citeauthoryear{{Schaye}, {Theuns}, {Leonard} \&
  {Efstathiou}}{{Schaye} et~al.}{1999}]{sch99a}
{Schaye} J.,  {Theuns} T.,  {Leonard} A.,    {Efstathiou} G.,  1999, \mnras,
  310, 57, \eprint{astro-ph/9906271}

\bibitem[\protect\citeauthoryear{{Sembach}, {Tripp}, {Savage} \&
  {Richter}}{{Sembach} et~al.}{2004}]{sem04a}
{Sembach} K.~R.,  {Tripp} T.~M.,  {Savage} B.~D.,    {Richter} P.,  2004,
  \apjs, 155, 351, \eprint{astro-ph/0407549}

\bibitem[\protect\citeauthoryear{{Shull}, {Smith} \& {Danforth}}{{Shull}
  et~al.}{2011}]{shu11a}
{Shull} J.~M.,  {Smith} B.~D.,    {Danforth} C.~W.,  2011, ArXiv e-prints,
  \eprint{1112.2706}

\bibitem[\protect\citeauthoryear{{Smith}, {Hallman}, {Shull} \&
  {O'Shea}}{{Smith} et~al.}{2011}]{smi11a}
{Smith} B.~D.,  {Hallman} E.~J.,  {Shull} J.~M.,    {O'Shea} B.~W.,  2011,
  \apj, 731, 6, \eprint{1009.0261}

\bibitem[\protect\citeauthoryear{{Spergel}, {Bean}, {Dor{\'e}}, {Nolta},
  {Bennett}, {Dunkley}, {Hinshaw} \& {Jarosik}}{{Spergel}
  et~al.}{2007}]{spe07a}
{Spergel} D.~N.,  {Bean} R.,  {Dor{\'e}} O.,  {Nolta} M.~R.,  {Bennett} C.~L.,
  {Dunkley} J.,  {Hinshaw} G.,    {Jarosik} N. e.~a.,  2007, \apjs, 170, 377,
  \eprint{astro-ph/0603449}

\bibitem[\protect\citeauthoryear{{Springel}}{{Springel}}{2005}]{spr05b}
{Springel} V.,  2005, \mnras, 364, 1105, \eprint{astro-ph/0505010}

\bibitem[\protect\citeauthoryear{{Springel}, {White}, {Jenkins}, {Frenk},
  {Yoshida}, {Gao}, {Navarro}, {Thacker}, {Croton}, {Helly}, {Peacock}, {Cole},
  {Thomas}, {Couchman}, {Evrard}, {Colberg} \& {Pearce}}{{Springel}
  et~al.}{2005}]{spr05a}
{Springel} V.,  {White} S.~D.~M.,  {Jenkins} A.,  {Frenk} C.~S.,  {Yoshida} N.,
   {Gao} L.,  {Navarro} J.,  {Thacker} R.,  {Croton} D.,  {Helly} J.,
  {Peacock} J.~A.,  {Cole} S.,  {Thomas} P.,  {Couchman} H.,  {Evrard} A.,
  {Colberg} J.,    {Pearce} F.,  2005, \nat, 435, 629, \eprint{0504097}

\bibitem[\protect\citeauthoryear{{Sutherland} \& {Dopita}}{{Sutherland} \&
  {Dopita}}{1993}]{sut93a}
{Sutherland} R.~S.,  {Dopita} M.~A.,  1993, \apjs, 88, 253

\bibitem[\protect\citeauthoryear{{Tepper-Garc{\'{\i}}a}, {Richter}, {Schaye},
  {Booth}, {Dalla Vecchia} \& {Theuns}}{{Tepper-Garc{\'{\i}}a}
  et~al.}{2012}]{tep12a}
{Tepper-Garc{\'{\i}}a} T.,  {Richter} P.,  {Schaye} J.,  {Booth} C.~M.,  {Dalla
  Vecchia} C.,    {Theuns} T.,  2012, \mnras, 425, 1640, \eprint{1201.5641}

\bibitem[\protect\citeauthoryear{{Tepper-Garc{\'\i}a}, {Richter}, {Schaye},
  {Booth}, {Dalla Vecchia}, {Theuns} \& {Wiersma}}{{Tepper-Garc{\'\i}a}
  et~al.}{2011}]{tep11a}
{Tepper-Garc{\'\i}a} T.,  {Richter} P.,  {Schaye} J.,  {Booth} C.~M.,  {Dalla
  Vecchia} C.,  {Theuns} T.,    {Wiersma} R.~P.~C.,  2011, \mnras, 413, 190,
  \eprint{1007.2840}

\bibitem[\protect\citeauthoryear{{Thom} \& {Chen}}{{Thom} \&
  {Chen}}{2008a}]{tho08b}
{Thom} C.,  {Chen} H.-W.,  2008a, \apjs, 179, 37, \eprint{0801.2381}

\bibitem[\protect\citeauthoryear{{Thom} \& {Chen}}{{Thom} \&
  {Chen}}{2008b}]{tho08a}
{Thom} C.,  {Chen} H.-W.,  2008b, \apj, 683, 22, \eprint{0801.2380}

\bibitem[\protect\citeauthoryear{{Tilton}, {Danforth}, {Shull} \&
  {Ross}}{{Tilton} et~al.}{2012}]{til12a}
{Tilton} E.~M.,  {Danforth} C.~W.,  {Shull} J.~M.,    {Ross} T.~L.,  2012,
  ArXiv e-prints, \eprint{1204.3623}

\bibitem[\protect\citeauthoryear{{Tripp}, {Meiring}, {Prochaska}, {Willmer},
  {Howk}, {Werk}, {Jenkins}, {Bowen}, {Lehner}, {Sembach}, {Thom} \&
  {Tumlinson}}{{Tripp} et~al.}{2011}]{tri11a}
{Tripp} T.~M.,  {Meiring} J.~D.,  {Prochaska} J.~X.,  {Willmer} C.~N.~A.,
  {Howk} J.~C.,  {Werk} J.~K.,  {Jenkins} E.~B.,  {Bowen} D.~V.,  {Lehner} N.,
  {Sembach} K.~R.,  {Thom} C.,    {Tumlinson} J.,  2011, Science, 334, 952,
  \eprint{1111.3982}

\bibitem[\protect\citeauthoryear{Tripp, Savage,  \& Jenkins}{Tripp
  et~al.}{2000}]{tri00a}
Tripp T.~M.,  Savage B.~D.,     Jenkins E.~B.,  2000, The Astrophysical Journal
  Letters, 534, L1

\bibitem[\protect\citeauthoryear{{Tripp}, {Sembach}, {Bowen}, {Savage},
  {Jenkins}, {Lehner} \& {Richter}}{{Tripp} et~al.}{2008}]{tri08b}
{Tripp} T.~M.,  {Sembach} K.~R.,  {Bowen} D.~V.,  {Savage} B.~D.,  {Jenkins}
  E.~B.,  {Lehner} N.,    {Richter} P.,  2008, \apjs, 177, 39,
  \eprint{0706.1214}

\bibitem[\protect\citeauthoryear{{Verner}, {Barthel} \& {Tytler}}{{Verner}
  et~al.}{1994}]{ver94a}
{Verner} D.~A.,  {Barthel} P.~D.,    {Tytler} D.,  1994, \aaps, 108, 287

\bibitem[\protect\citeauthoryear{{Weinberg}, {Miralda-Escude}, {Hernquist} \&
  {Katz}}{{Weinberg} et~al.}{1997}]{wei97b}
{Weinberg} D.~H.,  {Miralda-Escude} J.,  {Hernquist} L.,    {Katz} N.,  1997,
  \apj, 490, 564, \eprint{astro-ph/9701012}

\bibitem[\protect\citeauthoryear{{Wiersma}, {Schaye} \& {Smith}}{{Wiersma}
  et~al.}{2009a}]{wie09a}
{Wiersma} R.~P.~C.,  {Schaye} J.,    {Smith} B.~D.,  2009a, \mnras, 393, 99,
  \eprint{0807.3748}

\bibitem[\protect\citeauthoryear{{Wiersma}, {Schaye} \& {Theuns}}{{Wiersma}
  et~al.}{2011}]{wie11a}
{Wiersma} R.~P.~C.,  {Schaye} J.,    {Theuns} T.,  2011, \mnras, 415, 353,
  \eprint{1101.3550}

\bibitem[\protect\citeauthoryear{{Wiersma}, {Schaye}, {Theuns}, {Dalla Vecchia}
  \& {Tornatore}}{{Wiersma} et~al.}{2009b}]{wie09b}
{Wiersma} R.~P.~C.,  {Schaye} J.,  {Theuns} T.,  {Dalla Vecchia} C.,
  {Tornatore} L.,  2009b, \mnras, 399, 574, \eprint{0902.1535}

\bibitem[\protect\citeauthoryear{{Williger}, {Heap}, {Weymann}, {Dav{\'e}},
  {Ellingson}, {Carswell}, {Tripp} \& {Jenkins}}{{Williger}
  et~al.}{2006}]{wil06a}
{Williger} G.~M.,  {Heap} S.~R.,  {Weymann} R.~J.,  {Dav{\'e}} R.,  {Ellingson}
  E.,  {Carswell} R.~F.,  {Tripp} T.~M.,    {Jenkins} E.~B.,  2006, \apj, 636,
  631, \eprint{0505586}

\end{thebibliography}

 % used this when submitting both to journal and arXiv; be sure to include the .bbl file!

%--------------------------------------------------------------------------------------------------------------------------------------------------------------------------------

%--------------------------------------------------------------------------------------------------------------------------------------------------------------------------------
\label{lastpage}
%--------------------------------------------------------------------------------------------------------------------------------------------------------------------------------

\clearpage %to force LaTeX to output all floating objects processed before this line

%--------------------------------------------------------------------------------------------------------------------------------------------------------------------------------
% Appendix
\appendix

%--------------------------------------------------------------------------------------------------------------------------------------------------------------------------------
\section{Current status of \NeVIII\ observations at low redshift} \label{sec:current}

In this section we give a summary of the {\em intervening} \NeVIII\ absorbers detected at low redshift and reported in the literature to date. So-called 'proximate'  \NeVIII\ absorbers, which are believed to originate in the outflows of QSOs, have been also identified at low redshift \citep[e.g.][]{muz12a}, but they are out of the scope of this study.\\

The first \NeVIII\ absorber (with a total column density $\log_{10} (\NNeVIII / \psc ) = 13.85$ and corresponding rest-frame equivalent width $W_r \approx 33~$m\AA) was discovered by \citet{sav05a} at $z=0.207$ along the sightline to the QSO HE 0226-4110 ($z_{\rm em} = 0.495$) in a combined FUSE / STIS spectrum with S/N = 10 -- 20 per resolution element. Absorption by \HI\ and low ionisation states of other elements (e.g.\ \CIII, \NIII, \SiIII) and highly ionised oxygen (\OVI) where also  detected at the same redshift. While the low ionisation states could be explained by photo-ionisation, the same model failed to explain the observed \OVI\ - \NeVIII\ absorption. The implied path length ($\sim 10 ~$Mpc), in particular, would result in a Hubble broadening which is 10 times larger than the observed line width. \citet{sav05a} thus concluded that the \OVI\ - \NeVIII\ absorption is produced by collisionally ionised gas at a temperature $\log_{10} ( T / \K) \approx 5.7$. They estimated a total hydrogen column density $\log_{10} (\NH / \psc) \approx 20$ for this gas phase, suggesting that \NeVIII\ absorbers may harbour a significant fraction of the baryons in the Universe at low redshift. This absorption system was re-observed by \cite{sav11b} in a COS spectrum of the source HE 0226-4110 with S/N=20 -- 40. With the \NeVIII\ absorption out of the COS spectral range at this redshift, this group focused on the detection of a feature identified as a broad \lya\ absorber (BLA) with $\log_{10} (\NHI / \psc ) \approx 13.9$ and a Doppler parameter $\bneviii = 72^{+13}_{-6} \kms$, consistent with gas at a temperature in the range $\log_{10}( T / \K) = 5.4 - 5.7$. A joint model of the BLA - \OVI\ - \NeVIII\ system allowed them to constrain the temperature to $\log_{10}( T / \K) = 5.68 \pm 0.02$ and the total hydrogen column density to $\log_{10} ( \NH / \psc) = 20.06 \pm 0.09$, thus reinforcing the conclusion by \citet{sav05a} that \NeVIII\ absorbers at low redshift contain a significant amount of baryons.

A second \NeVIII\ detection ($\log_{10} (\NNeVIII / \psc ) = 13.98$, $W_r = 47~$m\AA) at $z = 0.326$ was reported by \citet{nar09b}. This absorber was identified in a FUSE spectrum of the QSO 3C 263 ($z_{\rm em} = 0.646$) with S/N = 5 - 10. This group found that the ionisation state of the gas phase giving rise to the observed \NeVIII\ absorption could not be explained by photo-ionisation. Rather, this absorber appears to be multi-phase, with the \NeVIII\ absorption arising in gas at $T \sim 10^6 \K$.  These authors provided the first estimate of the baryonic content of \NeVIII\ systems, $\Omega_b(\NeVIII) = 0.00263$, which corresponds to roughly 6 per cent of the total baryons in the Universe (assuming $\Omega_b = 0.0418$). This absorber was later confirmed in a COS spectrum by \citet{nar12a}. They re-analysed the physical conditions in this \OVI - \NeVIII\ absorbing gas, and found that is is best modelled with gas in CIE at $T = 5.2 \times 10^5 \K$ with $[\Ne / \H] = -0.12^{+0.12}_{-0.18}$ (adopting $\Xne_{\odot} = -4.07$). These authors noted for the first time the surprisingly small number of \NeVIII\ detections at low redshift\footnote{For comparison, there are more than 100 \OVI\ detections at low redshift reported in the literature to date \citep[see e.g.][]{til12a}.}. This fact led to the interpretation that the \NeVIII\ absorbers have a physical origin different from the filamentary structures of the highly ionised gas outside the virial boundaries of galaxies. This group suggested that the observed \NeVIII\ absorption originates in the extended regions around galaxies, i.e.\ the circum-galactic medium (CGM), which is consistent with the relatively high metal abundances of the \NeVIII\ absorbers.

\citet{nar11a} next reported on the detection of absorption by \NeVIII\ (and associated \OVI) in an intervening, multi-phase system at $z=0.495$ in a COS spectrum (S/N=15 - 20) along the \los\ to the source PKS 0405-123. Using the apparent optical depth method \citep[AOD; ][]{sav91a}, they measured $\log_{10} (\NNeVIII / \psc ) = 13.96$ and $W_r = 45~$m\AA. Allowing for the observed \OVI\ to have a significant contribution from the photo-ionised, cooler gas phase, the ionisation state of the warmer, \OVI - \NeVIII\ absorbing phase can be explained with a model that takes into account both photo- and collisional ionisation. Adopting a fixed temperature $T = 4.7 \times 10^5 \K$ and $[\Ne/\H] = -0.3$ (relative to $\Xne_{\odot} = -4.07$), \citet{nar11a} estimated a gas density $\nH \approx 8\times10^{-4} \pcc$, and a linear size of the absorber of roughly 20~kpc.

Another prominent example is the system in a COS spectrum of the QSO PG 1206+459 reported by \citet{tri11a}. This group detected a strong, complex \NeVIII\ absorber (and absorption by a series of other high and low ionisation states of other elements: \HI; \MgII, \MgX; \NII, \NIII, \NIV, \NV; \OIII, \OIV; \SuIII, \SuIV, \SuV) at $z = 0.927$, consisting of nine individual components with a total column density  $\log_{10} (\NNeVIII / \psc ) \approx 14.9$. They found strong evidence that the \NeVIII\ absorption is produced by the highly enriched, kinematically complex plasma in a strong outflow from a post-starburst galaxy ($L = 1.8~L^*$) at $z = 0.927$ and a projected impact parameter $\rho = 68~\kpc$ from the \los. They estimated that the warm-hot ($T > 10^5 \K$) gas phase contains 10 -- 150 times more mass than the cool ($T < 10^5 \K$) photo-ionised gas phase, with individual absorbing components harbouring $10 \time 10^8 ~\Msun$ -- $4 \times 10^{10} ~\Msun$.

Finally, three intervening systems at $z = 0.684$, 0.705,  and 0.725 along a single \los\ to the QSO PG 1148+549 containing \NeVIII\ (and \OVI) were   reported by \citet{mei13a}, These absorbers were detected in a COS spectrum with S/N = 20 - 40. Two of the absorbers ($z = 0.684$, $z=0.705$) appear to be two-component systems, with total $\log_{10} (\NNeVIII / \psc ) = 13.95$ and $W_r \approx 43~$m\AA, and total $\log_{10} (\NNeVIII / \psc ) = 13.86$ and $W_r \approx 35~$m\AA, respectively. The system at $z = 0.725$ shows a very simple kinematic structure (probably single-component) with $\log_{10} (\NNeVIII / \psc ) = 13.81$ and $W_r \approx 33~$m\AA. All three \OVI\ - \NeVIII\ absorbers can be explained as being produced in gas in CIE with a temperature around $\log_{10} ( T / \K) = 5.7$, and a metallicity (i.e.\ oxygen abundance) $[\Ox / \H] > -0.5$. This group  estimated that the density of gas as traced by these absorbers is $\Omega_b = 0.002$, or roughly 4 per cent of the baryon density (assuming $\Omega_b = 0.0418$). Given the small redshift path, these three detections imply a relatively high number density $dN/dz(W > 30 {\rm ~m\AA}) = 7^{+7}_{-4}$. However, their strong clustering along the \los\ appears unusual and these authors suggest that these three absorbers could result from this \los\ piercing a region with a unusual high density of star-forming galaxies. Hence, the high number density of \NeVIII\ absorbers inferred from this study might not be representative and likely overestimated as a result of cosmic variance.\\

%--------------------------------------------------------------------------------------------------------------------------------------------------------------------------------
% FIGURE: title
\begin{figure*}
\resizebox{0.33\textwidth}{!}{\includegraphics{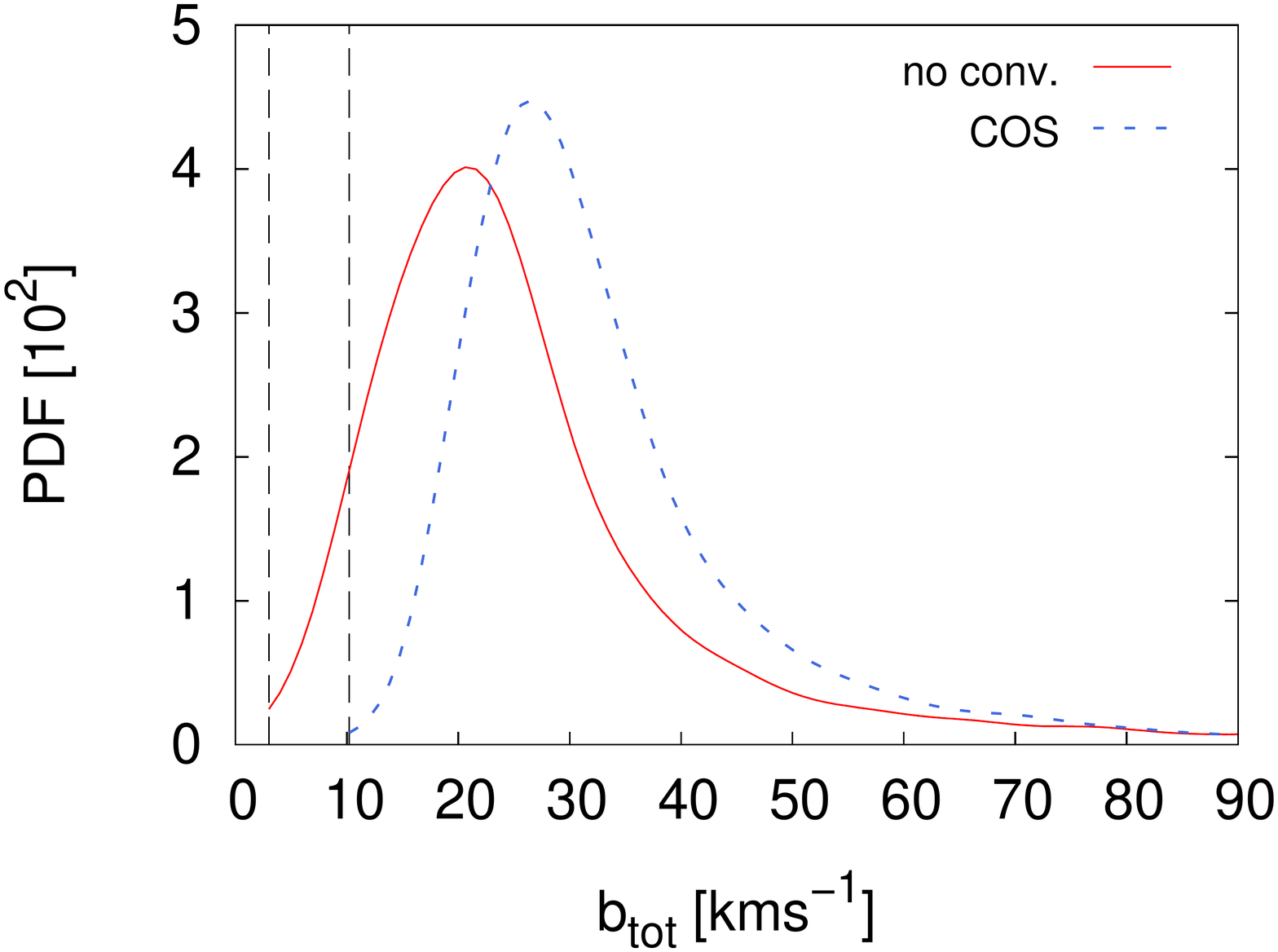}}
{\resizebox{0.33\textwidth}{!}{\includegraphics{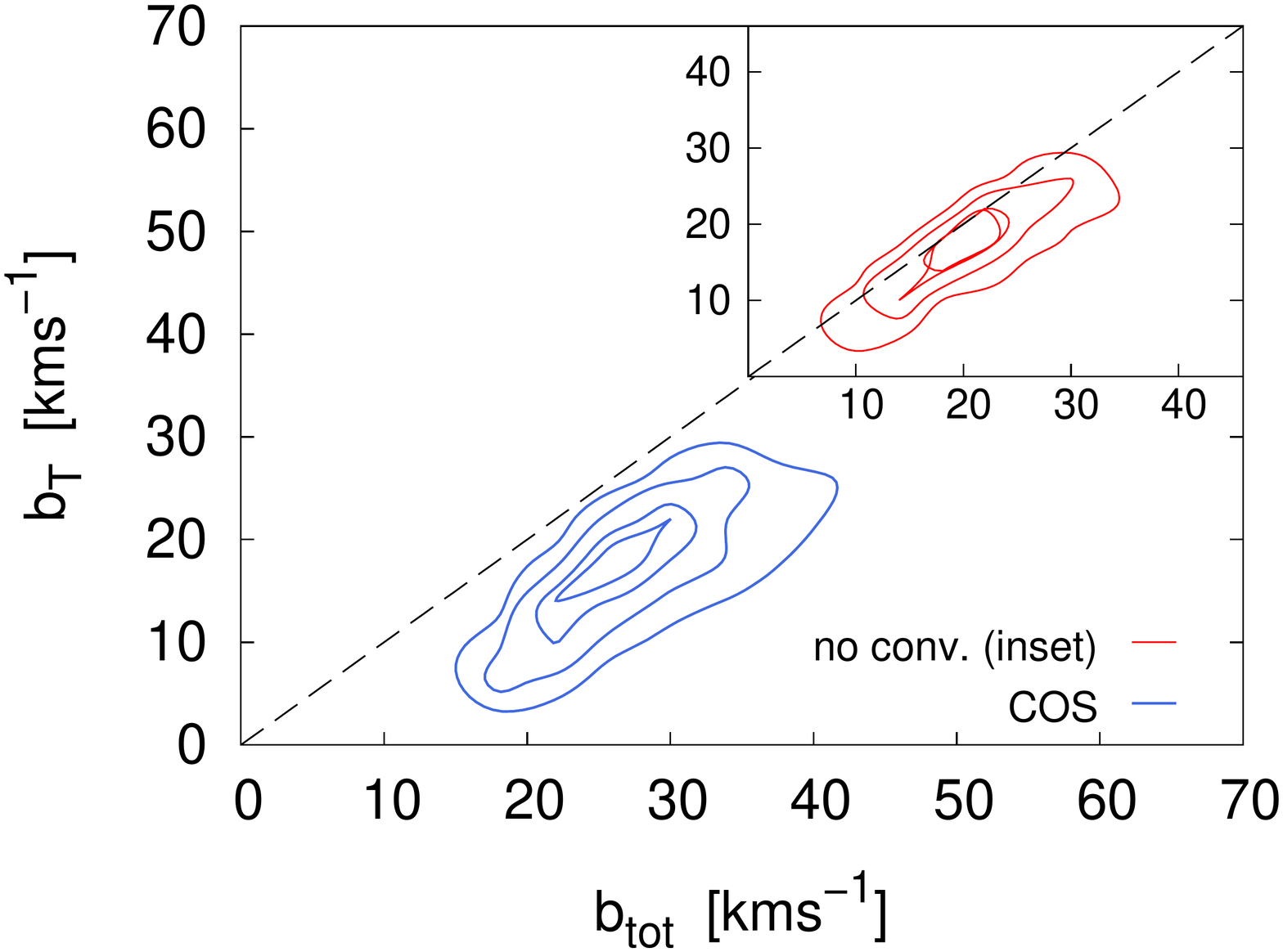}}}
{\resizebox{0.33\textwidth}{!}{\includegraphics{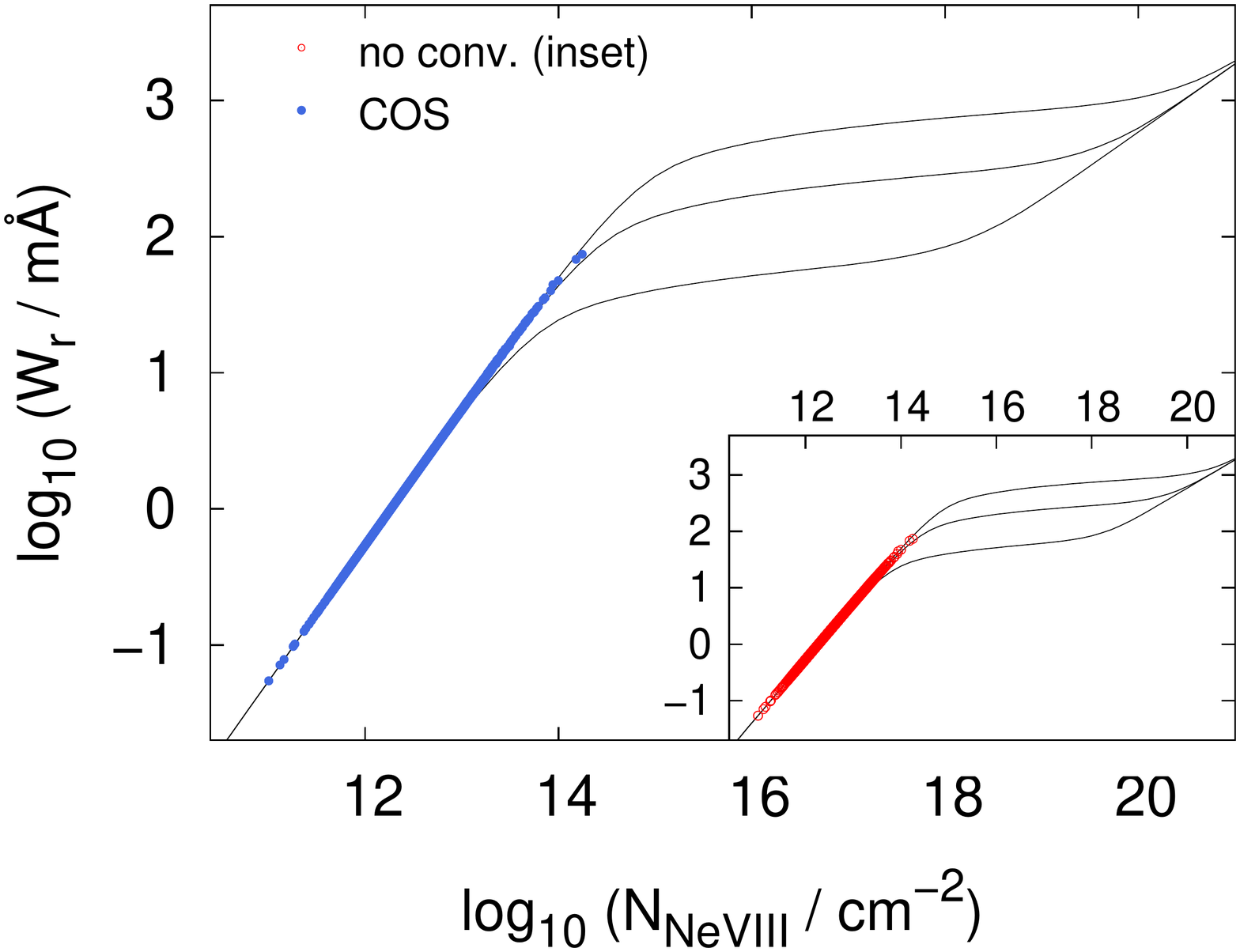}}}
\caption{{\em Left:} Distribution of Doppler parameters of the \NeVIII\ components. The dashed vertical line indicates in each case the resolution limit, i.e.\ the minimum adopted $b$-value. The blue dashed line shows the distribution of components identified in spectra that have been convolved with a Gaussian LSF ($FWHM = 17 \kms$), while the red solid line shows the distribution of components identified in spectra that have {\em not} been convolved. {\em Centre:}  Comparison of the thermal line width, $b_T$ (equation \ref{eq:bneviii}), to the total (fitted) line width, $\bneviii$, of the {\em single-component} \NeVIII\ absorbers identified in high-S/N spectra in our simulation at $z=0.5$. Clearly, the intrinsic width of the absorption lines is mainly set by temperature of the absorbing gas. However, convolution with an instrumental LSF results in absorption line widths that are significantly  larger than the thermal line width. Note that roughly half (45 per cent) and one third (35 per cent) of the \NeVIII\ absorbers appear as single-component features in the convolved and non-convolved spectra, respectively. {\em Right:} Rest-frame equivalent width and column density for our sample of \NeVIII\ absorbing components identified in 5000 spectra along random \los\ drawn from our simulation at $z=0.5$; each of the black curves shows a theoretical curve-of-growth (CoG) adopting $\bneviii = 50 \kms$, $18 \kms$, and $4 \kms$ (from top to bottom). Note that there is a linear relationship between rest-frame equivalent width and column density for all \NeVIII\ components. Hence, convolution with an instrumental LSF does not affect the measured column density in our spectra, since it conserves the equivalent width.}
\label{fig:cog}
\end{figure*}
%--------------------------------------------------------------------------------------------------------------------------------------------------------------------------------

%--------------------------------------------------------------------------------------------------------------------------------------------------------------------------------
\section{Synthetic spectra} \label{sec:spec}

The \NeVIII\ line statistics (Section \ref{sec:stats}) and the \NeVIII- and \HI-optical depth weighted quantities used throughout this work have been extracted from a set of spectra along 5000 random \loss\ drawn from our simulation at $z=0.5$, covering a total redshift path \mbox{$\Delta z = 212$}, corresponding to an absorption path length \mbox{$\Delta \chi = 387$}. For each \los\ we have computed a pair of synthetic spectra, with each spectrum containing absorption by \HI\ \lya\ or \NeVIII\ \NeVIIIstrong\ only. These spectra have been generated using the package \textsc{specwizard} written by Joop Schaye, Craig M.~Booth, and Tom Theuns, following the method described in detail in \citet[][ their Section 3.1]{tep11a}. We have convolved our spectra with a Gaussian line spread function (LSF) with a \mbox{FWHM = $17 \kms$}, and have re-sampled our spectra onto pixels of $3\kms$ in size, which toughly match the instrumental properties of COS. Finally, we have added Gaussian noise with a flux dependent root-mean-square (rms) amplitude given by \mbox{$\left({\rm S/N } \right)^{-1} F(v)$}, where $F(v)$ is the flux as a function of velocity along the \los, and S/N is the adopted signal-to-noise ratio; we adopt S/N=1000 to allow for the detection of most of the \NeVIII\ in our simulation (see below and Section \ref{sec:model}). Throughout this work, we shall refer to these as 'high-S/N, COS-resolution' spectra.

We fit our synthetic spectra using a significantly modified version of \textsc{autovp} \citep{dav97a} following the method detailed in \citet[][ their Appendix A]{tep12a}. Briefly, \NeVIII\ absorption lines are identified in regions with an (integrated) rest-frame equivalent width \mbox{$W_r \geq N ~\sigma_{W_r}$}, where \mbox{$\sigma_{W_r}$} is the uncertainty in the equivalent width, integrated over $n$ pixels, and $N$ is the significance level. We adopt $n=25$ (equivalent to $75 \kms$ or approximately four times our adopted spectral resolution) and\footnote{The \NeVIII\ detections reported to date in the literature have a significance $N\sigma \gtrsim 4 \sigma$.} $N = 5$. With these values, the significance value chosen translates into a formal, rest-frame equivalent width limit $W_r \approx 193 \left({\rm S/N}\right)^{-1} ~{\rm m\AA}$. This implies a {\em formal} completeness limit $W_r \approx 0.2~$m\AA\ or $\log_{10} (\NNeVIII / \psc) \approx 11.5$ for our spectra with S/N=1000. To ensure the detection of all relevant features, we impose a minimum column density $\log_{10} (\NNeVIII / \psc) = 11.0$, which is a factor 3 below the formal completeness limit. We note that the effective completeness limit is somewhat higher, $\log_{10} (\NNeVIII / \psc) \approx 11.7$ or $W_r \approx 0.3~$m\AA.

Each component is fitted assuming a Gaussian line profile,\footnote{The column density of the lines is low enough for the difference between a Gaussian and a Voigt line profile to be negligible.} and the parameters (velocity centroid $v_0$, column density $\NNeVIII$, and Doppler parameter $\bneviii$) of all lines identified in each region are simultaneously and iteratively adjusted using the Levenberg-Marquardt algorithm \citep{lev44a,mar63a} as implemented in \citet{num92a} until the reduced $\chi^2$-value, i.e.\ the $\chi^2$-value divided by the degrees of freedom, is below \mbox{$\chi^2_{\rm th} \equiv 1.25$}. During each iteration, lines with relative errors in $\NNeVIII$ or $\bneviii$ larger than 50 per cent are discarded. Note that each region is fitted with the {\em least number of absorption components} consistent with the condition $\chi^2 \leq \chi^2_{\rm th}$. Fitting of our 5000 simulated spectra yields a total of 8665 \NeVIII\ components, with roughly half (45 per cent) of the identified \NeVIII\ absorbers being single-component features. The total cumulative line frequency per unit redshift is $dN/dz(W_r > 0.3~{\rm m\AA}) \approx 41$.

To investigate the effect of the instrumental broadening on the line properties, we fit a second set of spectra which are identical to our high-SN spectra except that these new spectra are not convolved with an instrumental LSF. In Figure \ref{fig:cog} we compare the line parameters of the two \NeVIII\ absorption lines samples obtained from these two different sets. The left panel demonstrates that convolution with a Gaussian LSF artificially increases the observed line width by a significant amount, as expected. Therefore, while the intrinsic line width is a on average good indicator of the gas temperature (central panel, inset), the width of lines identified in convolved spectra can at most provide an upper limit. The right panel shows the relationship  between rest-frame equivalent width and column density (i.e.\ the curve-of-growth). Clearly, the equivalent width of the majority of the \NeVIII\ absorbers depends linearly on the \NeVIII\ column density. Since the convolution conserves the rest-frame equivalent width, the column density of the lines should be conserved as well.

%--------------------------------DOCUMENT END-----------------------------------
\end{document}